\documentclass[traditabstract]{aa} 

\usepackage[colorlinks=true, linkcolor=blue, citecolor=blue, urlcolor=blue]{hyperref}
\usepackage{amsmath}
\usepackage{graphicx}
\usepackage{txfonts}
\usepackage{amssymb}
\usepackage[]{natbib}
\usepackage{mathrsfs}
\usepackage{caption}
\usepackage{float}
\usepackage{afterpage}
\usepackage{amstext}

\begin{document}

   \title{TESS unveils the optical phase curve of KELT-1b}
   \subtitle{Thermal emission and ellipsoidal variation from the brown dwarf companion along with the stellar activity}
 \author{C. von Essen\inst{1,2}
          \and
          M. Mallonn\inst{3}
          \and
          A. Piette\inst{4}
          \and
          N. B. Cowan\inst{5}
          \and
          N. Madhusudhan\inst{4}
          \and
          E. Agol\inst{6}
          \and
          V. Antoci\inst{7, 1}
          \and
          K. Poppenhaeger\inst{3}
          \and
          K.G. Stassun\inst{8}
          \and
          S. Khalafinejad\inst{9}
          \and
          G. Tautvai{\v s}ien{\.e}\inst{2}
          }

   \institute{Stellar Astrophysics Centre, Department of Physics and Astronomy, Aarhus University, Ny Munkegade 120, DK-8000 Aarhus C, Denmark\\
         \email{cessen@phys.au.dk}
         \and
             Astronomical Observatory, Institute of Theoretical Physics and Astronomy, Vilnius University, Sauletekio av. 3, 10257, Vilnius, Lithuania
        \and
             Leibniz-Institut f\"{u}r Astrophysik Potsdam (AIP), An der Sternwarte 16, D-14482 Potsdam, Germany
        \and
             Institute of Astronomy, University of Cambridge, Madingley Road, Cambridge, CB3 0HA, UK
        \and 
             Department of Physics and Department of Earth \& Planetary Sciences, McGill University, Montr\'{e}al, QC H3A 2T8, Canada
        \and        
             Department of Astronomy and Virtual Planetary Laboratory, University of Washington, Seattle, WA 98195, USA
        \and 
            DTU Space, National Space Institute, Technical University of Denmark, Elektrovej 328, DK-2800 Kgs. Lyngby, Denmark 
        \and
            Vanderbilt University, Department of Physics \& Astronomy, 6301 Stevenson Center Ln., Nashville, TN 37235, USA
        \and
            Landessternwarte, Zentrum f\"{u}r Astronomie der Universit\"{a}t Heidelberg, K\"{o}nigstuhl 12, 69117 Heidelberg, Germany
             }
   \date{Received ----; accepted ----}
   
\abstract{We present the detection and analysis of the phase curve of KELT-1b at optical wavelengths, analyzing data taken by the Transiting Exoplanet Survey Satellite (TESS) during cycle 2 and sector 17. With a mass of $\sim$27 $M_{\rm Jup}$, KELT-1b is an example of a low-mass brown dwarf. Due to the high mass and close proximity of its companion, the host star exhibits a TESS light curve that shows clear ellipsoidal variations. We modeled the data with a six-component model: secondary eclipse, phase curve accounting for reflected light and thermal emission, Doppler beaming, ellipsoidal variations, stellar activity, and the primary transit. We determined the secondary eclipse depth in the TESS bandpass to be \mbox{304 $\pm$ 75} parts-per-million (ppm). In addition, we measured the amplitude of the phase curve to be \mbox{128 $\pm$ 27 ppm}, with a corresponding eastward offset between the region of maximum brightness and the substellar point of 19.2 $\pm$ 9.6 degrees, with the latter showing good agreement with Spitzer measurements. We determined a day-side brightness temperature in the TESS bandpass of \mbox{3201 $\pm$ 147 K} that is approximately 200~K higher than the values determined from the Spitzer 3.6 and 4.5 $\mu$m data. By combining TESS and Spitzer eclipse depths, we derived a day-side effective temperature of \mbox{$T_\mathrm{eff}$ = 3010 $\pm$ 78 K}. Previously published eclipse depths in the near-infrared suggest a much higher brightness temperature and this discrepancy cannot be explained by spectral models combined with the current data. We attribute those large eclipse depths to unmodeled ellipsoidal variations, which would typically be manifested as a deeper secondary eclipse in observations with insufficient phase coverage. A one-dimensional self-consistent atmospheric model is able to explain the TESS and Spitzer day-side brightness temperatures with thermal emission alone and no reflected light. The difference between the TESS and Spitzer brightness temperatures can be explained via CO absorption due to a non-inverted temperature profile. The night side data fix an upper limit of $\sim$2000~K on the internal temperature of KELT-1~b. }


\maketitle

\section{Introduction}

A brown dwarf that transits its host star once per orbital period offers a unique opportunity to characterize its atmosphere. The signal of the object can be separated from the light of the host star by time-differential techniques \citep{Charbonneau2007}, that is, by carrying out a comparison of the brightness of the star-companion system based on either: out-of transit with in-transit or out-of secondary eclipse with in-eclipse. The wavelength-resolved application of these techniques is known as transmission spectroscopy for the transit event and emission spectroscopy for the secondary eclipse observation \citep[see reviews by][]{Kreidberg2018b,Deming2019,Madhusudhan2019}. In recent years, these methods have been applied very successfully for close-in planets \citep[e.g.,][]{Sing2016,Evans2016,Tsiaras2019}  because of the more frequent occultations and  also because hotter planets have larger atmospheric scale heights and greater thermal emission.

Close-in companions are generally expected to be tidally locked as the tidal synchronization timescale is typically much shorter than the estimated stellar age \citep{Guillot1996}. Thus, they possess permanent day and night sides.  The two successful techniques --- transmission and emission spectroscopy --- probe different regions: the day-night terminator and the day side, respectively, which are separated in longitude and exhibit different physical conditions. One way to observe the dependence of atmospheric parameters (such as temperature) on the planetary longitude is the observation of a full-orbit phase curve \citep[see recent reviews by][]{Shporer2017,Parmentier2018}. At infrared wavelengths, the temperature difference between the day side and night side causes a brightness variation modulated with the planetary rotation due to varying thermal emission. At optical wavelengths, the brightness modulation is a combination of thermally emitted light and reflected stellar light, both having a maximum around secondary eclipse and a minimum around the transit event. However, typical amplitudes of phase curve modulations are small, amounting to 100 ppm of the stellar light at optical wavelengths \citep{Esteves2015,Shporer2017,Shporer2019} and 1000~ppm in the near-infrared \citep{Beatty2014,Zhang2018,Kreidberg2018}, with the variations being very gradual over timescales of days. Thus, exoplanet phase curve observations are currently out of reach with regard to ground-based telescopes and can only be performed with space telescopes.

The Transiting Exoplanet Survey Satellite \citep[TESS,][]{Ricker2015} has measured the full-orbit phase curves of multiple close-in gas giants thus far, revealing the day- and night-side brightness temperatures at optical wavelengths as well as the phase curve offset between the brightest measured longitude and the most strongly irradiated longitude at the substellar point \citep[e.g.,][]{Shporer2019,Wong2019b,Wong2020,Wong2020b,Bourrier2019,Daylan2019,Jansen2020,Nielsen2020,vonEssen2020b}. These values provide information on the Bond albedo and the energy transport from the day to the night side \citep{Cowan2011}. Additionally, the TESS phase curves allowed for constraints on the geometric albedo of individual objects \citep{Shporer2019,Wong2020,Wong2020b,vonEssen2020b} and the detection of gravitational interactions between planet and host star \citep{Shporer2019,Wong2019b,Nielsen2020,Wong2020b,vonEssen2020b}, allowing for independent mass constrains. When compared with near-infrared phase curve measurements of the same objects, it is possible to study the wavelength dependence of the phase curve parameters and distinguish between several models for the pressure-temperature profile of the planet atmosphere \citep{Parmentier2018}. 

In this work, we present our analysis of the TESS phase curve of KELT-1b \citep{Siverd2012}. KELT-1b was discovered with the Kilodegree Extremely Little Telescope \citep[KELT,][]{Pepper2012} and was found to have a mass in the brown-dwarf regime, via subsequent radial-velocity measurements. The first measurements of the thermal emission of the day side were provided by \cite{Beatty2014} with secondary eclipse observations with the Spitzer space telescope at 3.6 and 4.5~$\mu$m and with ground-based observations in the Sloan z' band. A spectrally-resolved secondary eclipse was presented by \cite{Beatty2017} at the near-infrared H band, which showed a wavelength dependence of the brightness temperature of KELT-1b when compared to the previous Spitzer and z' band measurements. The day-side emission spectrum favored a model with monotonically decreasing temperature with altitude, similar to an isolated brown dwarf of the same temperature. Ultra-hot Jupiters of lower mass showed indications of isothermal pressure-temperature profiles \citep{Parmentier2018}, implying that the higher surface gravity of KELT-1b might play a major role in governing its atmospheric structure. Spitzer full-orbit phase curves at 3.6 and 4.5~$\mu$m were published by \cite{Beatty2019}, with values for heat redistribution efficiency similar to those of Jupiter-mass ultra-hot planets.

This work is structured as follows. In Section~\ref{sec:obs}, we present a detailed description of the observations. In Section~\ref{sec:analysis}, we present our analysis of the third light contamination, limb darkening coefficients, and primary transit parameters. We conclude that section by providing updated parameters for the host star and companion, which we use in our modeling. In Section~\ref{sec:modelling}, we detail the six different model components we take into consideration in order to characterize the atmospheric parameters of KELT-1b. In Section~\ref{sec:results}, we present the physical parameters derived from TESS observations and our analysis. In Section~\ref{sec:context} we set our observations into context. We present our concluding remarks in Section~\ref{sec:conclusion}.

\section{Observations and data preparation}
\label{sec:obs}

KELT-1 (TIC identifier 432549364) was observed by TESS between 8 and 31 of October 2019, during cycle 2 and sector 17, using camera 2. The photometric time series has a cadence of 120 seconds and it was constructed using the Science Processing Operations Center (SPOC) pipeline, based on the NASA Kepler mission pipeline \citep{Jenkins2016,Jenkins2017}. The light curve analyzed in this work is the one constructed by the  Presearch Data Conditioning (PDC) \mbox{msMAP} correction method. The timestamps are provided by SPOC in Barycentric Julian dates (TDB) and are, consequently, left unchanged. 

It is well known that the PDC data might show systematic features in the photometry that have not, in turn, been observed in the Aperture Photometry (SAP) data \citep{Bourrier2019,Shporer2019,Wong2020}. The unbinned KELT-1 photometry show variability during secondary eclipse and along the orbital phase. Consequently, to make a thoughtful choice between carrying out an unphysical detrending of the photometry (SAP), for instance, by detrending the data with a spline function or using the more physically-motivated detrending (PDC) for the cost of extra noise, we quantified the amount of correlated noise in the light curves computing and comparing their respective $\beta$ values \citep{Carter2009}. As was previously done by other authors \citep[see e.g.,][]{Bourrier2019,Shporer2019,Wong2020}, we first removed all data marked by a bad quality flag by the SPOC pipeline and then we ran a 16 point-wide moving median filter to the primary and secondary eclipse-masked light curve. We used this median filter to remove 3-$\sigma$ outliers. As was done by other authors, we removed the first 0.5 days of data from each TESS orbit \citep{Wong2020,Wong2019b}. The total number of points removed along the whole process was on the order of 25\%. This exercise was repeated over both the SAP and PDC light curves. Finally, we divided the light curve of each orbit by its median value, computed from the primary transit and secondary eclipse masked light curves. We did not carry out any other detrending nor normalization strategy \citep[see e.g.,][]{Huang2018,Wang2019} to avoid removing phase variations that are of relevance to our analysis. Instead, we compensate for any remaining noise by increasing the individual photometric error bars by the amount of correlated noise. 

To estimate the amount of correlated noise in both versions of the light curve SAP and PDC, we employed the method of residual binning described in \cite{Winn2008}. From the light curve, we subtracted our joint best-fit model, which is described in detail in Section~\ref{sec:modelling}. We then divided the light curve residuals into $M$ bins of equal duration, each one holding $N$ data points. If the data are not affected by correlated noise, they should follow the expectation of independent random numbers: 

\begin{equation}
  \hat{\sigma_N} = \sigma_1 N^{-1/2}[M/(M-1)]^{1/2}\ .
\end{equation}

\begin{figure*}[ht!]
    \centering
    \includegraphics[width=\textwidth]{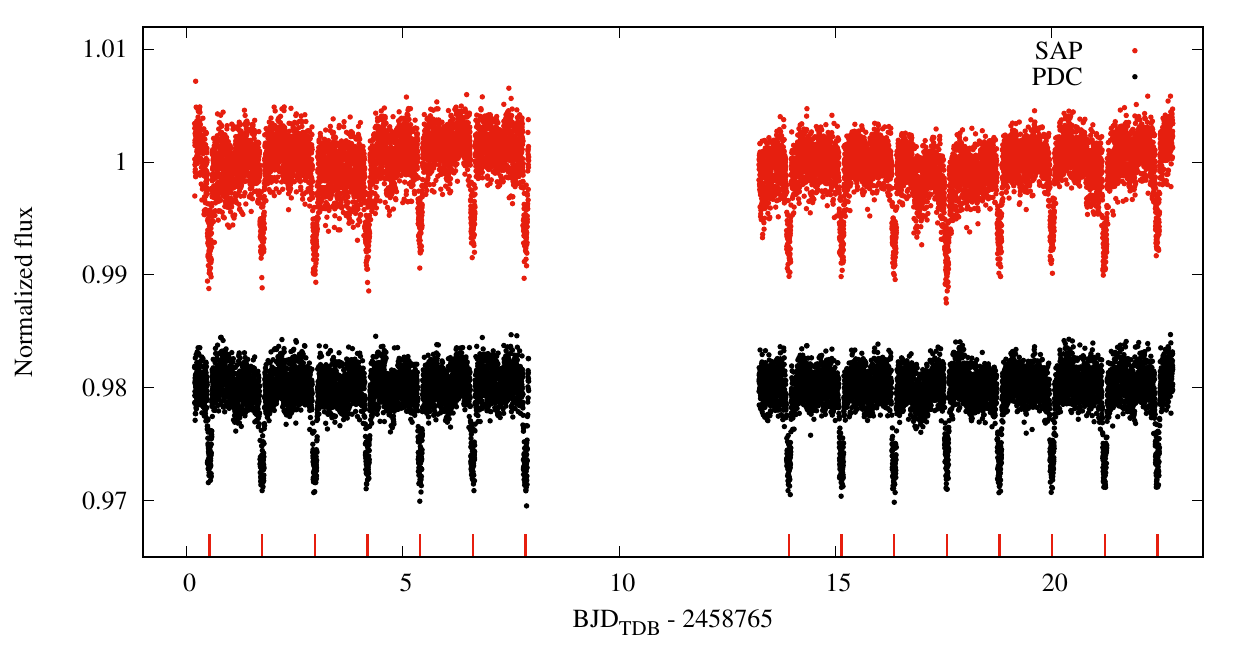}
    \caption{\label{fig:KELT1LC} Detrended flux of KELT-1 observed by TESS. SAP photometry is indicated with red points and PDC with black circles. Both light curves are shown after our internal data preparation. The fifteen transits analyzed in this work are indicated in the bottom of the figure with red lines.}
\end{figure*}

\noindent In the equation, $\sigma_1^2$ is the sample variance of the unbinned data, and $\hat{\sigma_N^2}$ is the sample variance (or RMS) of the binned data:

\begin{equation}
  \sigma_N = \sqrt{\frac{1}{M}\sum_{i = 1}^{M}(<\hat{\mu_i}> - \, \hat{\mu_i})^2},
\end{equation}

\noindent where $\hat{\mu_i}$ is the mean value of the residuals per bin and $<\hat{\mu_i}>$ is the mean value of the means. When correlated noise is present in the data, each $\sigma_N$ will differ by a factor of $\beta_N$ from their expectation, $\hat{\sigma_N}$. We may account for correlated noise by enlarging uncertainties by a factor $\beta$. We  compute the average of the $\beta_N$'s derived from different time bins relevant to the timescales involved in the study. For instance, when primary transits are fit, relevant timescales are usually close to the ingress or egress time \citep[see e.g.,][]{vonEssen2013}. 
In this work, the study spans the whole orbital phase of KELT-1b. As a consequence, we considered the timescale $\Delta t$ between a fourth of the transit duration to the whole orbital period, divided into steps of a fourth of the transit duration. This corresponds to a total of 84 $\Delta t$'s.  We then averaged all the $\beta_N$'s computed in this range to obtain the relevant $\beta$. Our results are \mbox{$\beta_{SAP}$ = 8.61} and \mbox{$\beta_{PDC}$ = 3.06}. These values do not come as a surprise, as the PDC data have their own detrending, while the SAP data do not. In any case, we adopt the PDC data, but with their individual photometric error bars enlarged by $\beta_{PDC}$ to alleviate the effects caused by correlated noise. We show the photometry of KELT-1 in Figure~\ref{fig:KELT1LC}, for both SAP (red points) and PDC (black points) light curves. TESS stared at the target for about 23 days, with the usual gap due to a data downlink. 

\section{Analysis and model considerations}
\label{sec:analysis}

\subsection{Third light contamination}
\label{sec:third_light}

Companions, whether they are visual or bound by gravity, dilute the depth of transits and must be properly accounted for. Due to TESS's large pixels projected on the sky (21$\times$21 arcsec each), it is very likely that flux from a neighbouring star will fall into the aperture around our target of interest.  PDC corrects the flux for each star falling into KELT-1's aperture. The amount of contaminated light found in the CROWDSAP keyword of KELT-1's fits file is 1.181012\%. Nonetheless, we carried out our own investigation to compare the results. 

In analyzing the target pixel file, we discovered one relatively bright visual companion included in the TESS aperture (\mbox{Gaia DR2 2881784379713578496}, henceforth, C1), and a second one that is much fainter (\mbox{Gaia DR2 2881784280929656448}, henceforth C2). According to Gaia's Data Release 2 \citep{GaiaCollaboration2018}, C1 has a magnitude of \mbox{G$_\mathrm{C1}$ = 15.2428 $\pm$ 0.0005} and an effective temperature of \mbox{$\sim$5300 K}, while C2 is about two magnitudes fainter, \mbox{G$_\mathrm{C2}$ = 17.3132 $\pm$ 0.0009}, and of unknown temperature. For comparison, KELT-1's Gaia magnitude is \mbox{$G_\mathrm{KELT-1}$ = 10.5905 $\pm$ 0.0003}. Additional targets have been found by GAIA DR2 with magnitudes $G>20$ and are, consequently, negligible. Furthermore, \cite{Siverd2012} found a companion $\sim$0.5 arcsec away from KELT-1, with a magnitude difference of \mbox{$\Delta$H = 5.90 $\pm$ 0.10}, thus, it is negligible as well at TESS wavelengths. Surprisingly, this target was not detected by Gaia. In addition to these, systematic surveys such as \cite{Mugrauer2019,Piskorz2015,Belokurov2020} do not list KELT-1 as having another close-in ($<$2 arcsec) companion. 

To compensate for the difference in distance between KELT-1 and C1, C2, we used Pogson's law:
\begin{equation}
m_\mathrm{KELT-1} - m_\mathrm{C1,C2} = -2.5 \times log_{10}\left(\frac{f_\mathrm{KELT-1}}{f_\mathrm{C1,C2} \times D} \right),
\end{equation}
where $m$ corresponds to the Gaia magnitudes, $f$ to the corresponding fluxes, and $D$ is a factor that compensates for the difference in distance. To compute the fluxes, $f$,  we made use of PHOENIX stellar intensities \citep{phoenix} convolved with Gaia's filter response. Solving this equation, we determined $D$ and we used it to scale down the spectra within the TESS transmission response. As previously mentioned, for C1 there are reliable measurements of the effective temperature of the star of \mbox{$\sim$5300 K}, so we used PHOENIX spectra matching this value. In the case of C1, the third light contribution ascends to 1.74\%. However, there is no information about the effective temperature of C2. As a consequence, we tried several PHOENIX spectra with effective temperatures ranging between 3500 and 8500 K, with a step of 1000 K. In each case, we computed the third light contribution in TESS bandpass. As extreme values, we found that for \mbox{T$_\mathrm{eff}$ = 8500 K} the third light contribution was 0.17\%, while for \mbox{T$_\mathrm{eff}$ = 3500 K} it increased to 0.28\%. As a consequence, the third light contribution lies between 1.91 and 2.02\%. These values are between 60 and 70\% larger than the value reported in CROWDSAP. We did not correct the third light contribution, but we pay special attention to the transit depth, as compared to the literature values. While the difference between 1.18\% and 2.0\% third light contribution might be measurable for the transit depth, its effect is negligible for the phase curve amplitude and the secondary eclipse depth.

\begin{figure}[ht!]
    \centering
    \includegraphics[width=.5\textwidth]{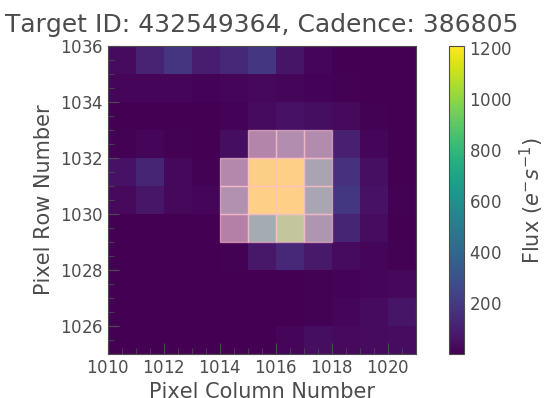}
    \includegraphics[width=.25\textwidth]{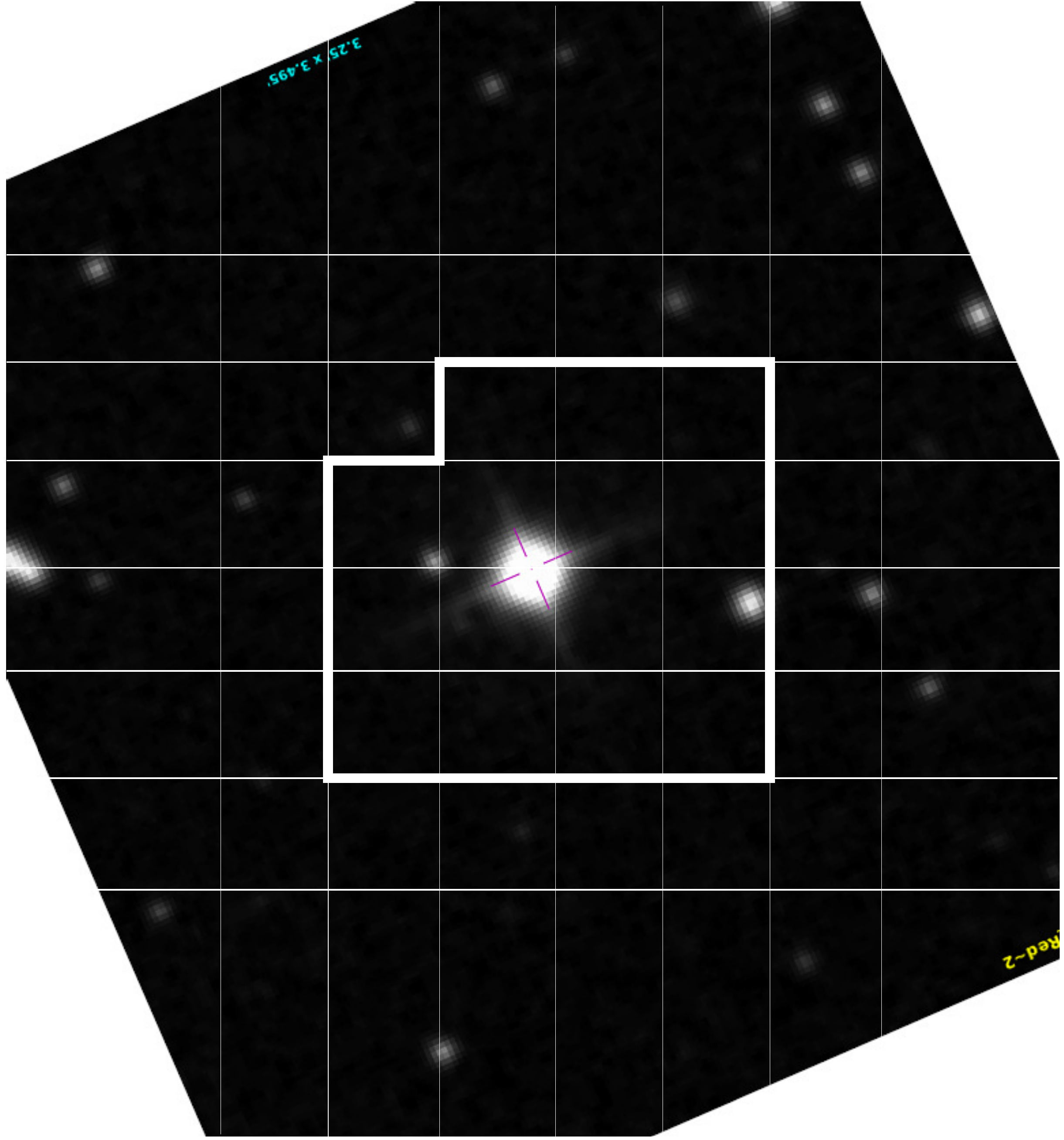}
    \caption{\label{fig:W33_aperture} {Third light contamination in the TESS aperture. \it Top:} Target Pixel File (TPF) of KELT-1 showing the aperture mask. Three stars are included and completely blurred. {\it Bottom:} $\sim$3$\times$3~arcmin field of view centered around KELT-1, the brightest star in the field. The mask and the pixels are schematized with white thick and thin lines, respectively. The field of view has been rotated to be aligned with the ecliptic.}
\end{figure}

\subsection{Limb-darkening coefficients}
\label{sec:LDCs}

To represent the stellar center-to-limb intensity variability we used a quadratic limb-darkening law:

\begin{equation}
\frac{I(\mu)}{I(1)} = 1 - u_1(1-\mu) - u_2(1-\mu)^2,
\label{eq_qua}
\end{equation}

\noindent with linear ($u_1$) and quadratic ($u_2$) limb-darkening coefficients (LDCs). In this equation, the intensity is normalized to that of the stellar center, and $\mu = \cos(\gamma)$, where $\gamma$ is the angle between the line of sight and the emergent intensity. To compute the limb-darkening coefficients, we made use of the TESS transmission response and angle-dependent, specific intensity spectra taken from PHOENIX library \citep{phoenix}. For the stellar parameters, we used an effective temperature of \mbox{$T_{\rm eff} = 6500$}~K, a surface gravity of \mbox{log($g$) = 4.5}, and a metallicity of \mbox{[Fe/H] = 0.00}, all fully consistent with KELT-1 values within errors. Similarly to \cite{vonessen2017} and \cite{claret2011}, we neglect the data points between \mbox{$\mu$ = 0} and \mbox{$\mu$ = 0.075}, as the intensity drop given by PHOENIX models is too steep and potentially unrealistic. After integrating the PHOENIX angle-dependent spectra convolved with the TESS response, we fit the derived intensities normalized by the maximum values with Equation~\ref{eq_qua} with a Markov-chain Monte Carlo (MCMC) approach. The derived limb-darkening coefficients for KELT-1 are \mbox{$u_1$ = 0.319(1)} and \mbox{$u_2$ = 0.227(3)}. Errors for the coefficients are derived from the posterior distributions of the MCMC chains after visually inspecting them for convergence. In order to assess the quality of our procedure, we fit the LDCs to TESS primary transit light curves. From their posterior distributions, we obtained consistent results with their PHOENIX counterparts. In order to reduce computational cost, in this work we consider $u_1$ and $u_2$ as fixed. To double check our procedure, we found our values comparable to those reported by \cite{Claret2017}, equal to \mbox{$u_1$ = 0.3079} and \mbox{$u_2$ = 0.2295}.

\subsection{Primary transit parameters derived from TESS light curves}
\label{sec:PT}

To derive updated transit parameters for KELT-1b for the purposes of our joint model as prior information, we fit TESS transits of KELT-1b. For this, we considered the data around primary transits only, plus and minus $\sim$1.5 hours of off-transit data to allow for a proper normalization. As a detrending model, we used a second-order time-dependent polynomial. The degree of the polynomial was determined from a prior fit to the data of a first, second, and third degree polynomials, after which we computed the Bayesian Information Criterion (BIC) from the residuals and analyzed which polynomial systematically minimized the BIC. All the transit light curves were fit simultaneously to a primary transit model \citep{MandelAgol2002}, while each primary transit light curve had its own set of detrending coefficients. To evaluate whether the non-zero eccentricity of the system, \mbox{e = 0.0013 $\pm$ 0.0005} \citep{Beatty2019}, has any impact on the derived primary transit parameters, we considered two orbits for KELT-1b, a circular one and an eccentric one, the latter with an eccentricity and argument of periastron fixed to the values specified in \cite{Beatty2019}. The transit photometry will not allow us to distinguish between the small-eccentricity and a zero-eccentricity model, but we aim to know whether the derived transit parameters differ. The fitting parameters for the primary transit model in the case of a circular orbit are the semi-major axis, $a/R_S$, the inclination in degrees, i, the orbital period, $P$, the planet-to-star radius ratio, $R_P/R_S$, and the mid-transit time of reference, $T_c$. As described in Section~\ref{sec:LDCs}, we considered a quadratic limb-darkening law with fixed limb-darkening coefficients. For a non-circular orbit the primary transit parameters are the semi-major axis, $a/R_S$, the inclination in degrees, $i$, the orbital period, $P$, the planet-to-star radius ratio, $R_P/R_S$, the time of periastron passage, $T_\mathrm{peri}$, the argument of periastron in degrees, $\omega$, and the orbital eccentricity, $e$. The latter two parameters $e$ and $\omega$ remained fixed throughout the fit, whereas the other were treated as free to fit.

\begin{figure}[ht!]
    \centering
    \includegraphics[width=0.5\textwidth]{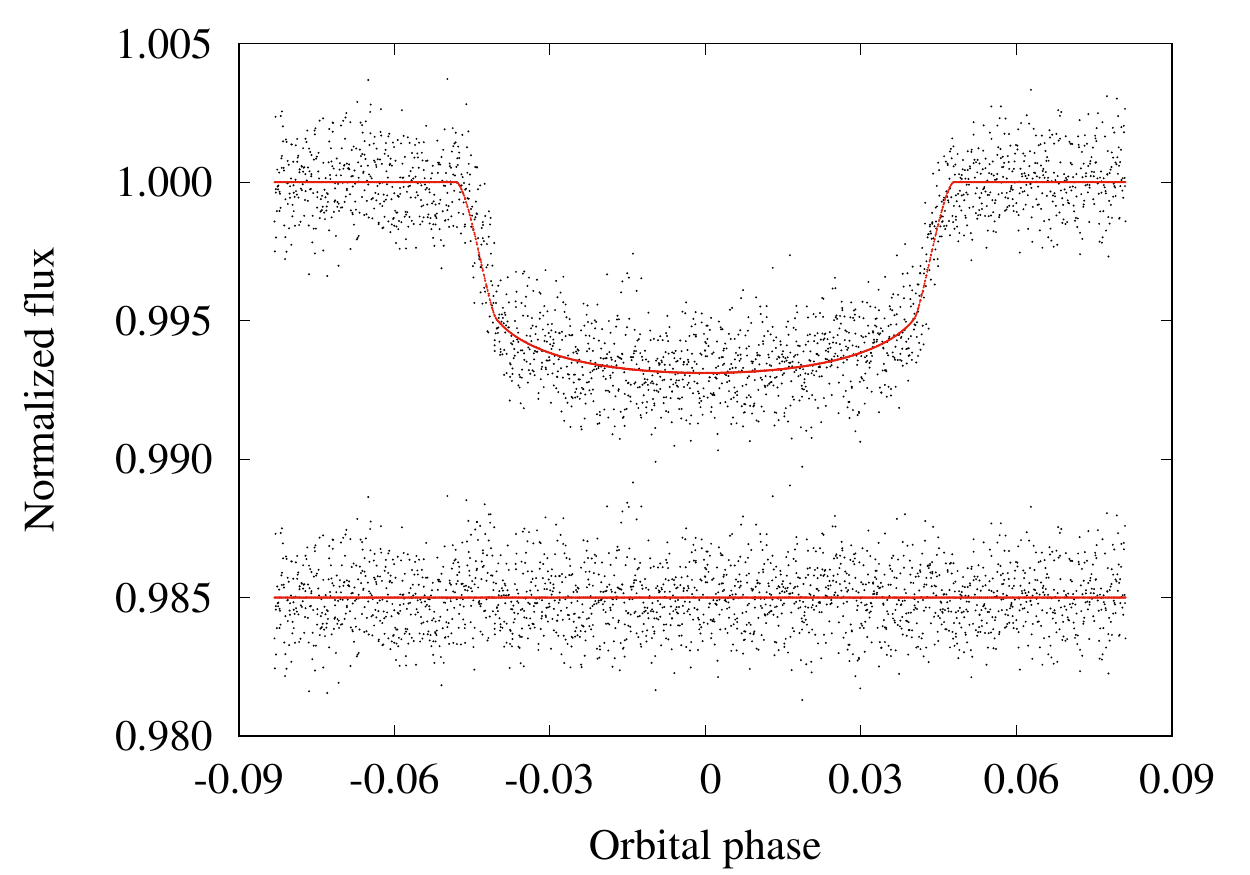}
    \caption{\label{fig:transits_folded} Primary transit light curve and residuals for KELT-1b for a circular orbit as a function of the orbital phase. Black points denote TESS data. Continuous red line corresponds to the best-fit primary transit model. The residuals have been artificially shifted below the transits.}
\end{figure}

To derive the best-fit values for the model parameters and their corresponding uncertainties we used a Markov-chain Monte Carlo (MCMC) approach, all implemented in routines of PyAstronomy\footnote{www.hs.uni-hamburg.de/DE/Ins/Per/Czesla/PyA/PyA/index.html} \citep{Patil2010,Jones2001}. We iterated $10^6$ times, with a conservative burn-in of the first 20\% of samples. For all the parameters we considered uniform priors around $\pm$50\% their corresponding starting values, which were, in turn, taken from \cite{Beatty2019}. These are specified in the last column of Table~\ref{tab:transit_parameters}. We computed the mean and standard deviation from the posterior distributions and used these as our best-fit values and 1-$\sigma$ uncertainties. We checked the convergence of the chains by visually inspecting each one of them and by dividing the chains in three sub-chains. In each case, we computed the usual statistics and we considered that a chain converged if the derived parameters were consistent within their uncertainties. The best-fit transit parameters for both the circular and non-circular solution is given in Table~\ref{tab:transit_parameters}. The posterior distributions and the corresponding correlations between parameters for the circular orbit can be seen in Appendix~\ref{fig:posterior_transits}. Besides the well known correlation between $a/R_S$ and $i$, and $P$ and $T_c$, the remaining parameters are uncorrelated, with Pearson's correlation values ranging between $-0.05$ and 0.05. The very small eccentricity indicated by \cite{Beatty2019} would not affect the derivation of the parameter values, we find all values to be consistent within 1\,$\sigma$ uncertainty. 
The phase-folded light curves for the circular orbit can be seen in Figure~\ref{fig:transits_folded}.

\begin{table*}[ht!]
    \centering
    \caption{\label{tab:transit_parameters} Best-fit transit parameters obtained from TESS photometry (this work), for a circular orbit (first column) and an eccentric orbit (second column), compared to those determined by \cite{Beatty2019}. When uncertainties are not given, it means that the parameter was considered as fixed. $T_c$ and $T_\mathrm{peri}$ are given -2450000 days.}
    \begin{tabular}{l c c c}
    \hline\hline
    Parameter        &    This work        &    This work            & \cite{Beatty2019}   \\
                     &  circular orbit     & eccentric orbit         & for comparison  \\
    \hline
    $a/R_S$          & 3.630 $\pm$ 0.051   &  3.652 $\pm$ 0.046      & 3.693 $\pm$ 0.038 \\
    $i$ ($^{\circ}$) & 87.2 $\pm$ 1.6      &  87.5 $\pm$ 1.4         & 86.8 $\pm$ 0.8 \\
    $R_P/R_S$        & 0.07688 $\pm$ 0.00040 & 0.07704 $\pm$ 0.00029 & 0.0771 $\pm$ 0.0003 \\
    $P$ (days)       & 1.2174928 $\pm$ 2.7$\times$10$^{-6}$   &   1.2174928 $\pm$ 3.3$\times$10$^{-6}$   &   1.2174928 $\pm$ 6$\times$10$^{-7}$ \\    $T_c$ (BJD$_\mathrm{TDB}$) & 8764.31647 $\pm$ 0.00018 &  -      & 7306.97602 $\pm$ 0.0003 \\
    $T_\mathrm{peri}$&   -                 & 8764.61506 $\pm$ 0.00028 &  - \\
    $\omega$ (deg)   &  270                & 358                     & 358 $\pm$ 51 \\      
    $e$              &  0                  & 0.0013                  & 0.0013 $\pm$ 0.0005 \\
    $\chi^2$         &  1566               &  1695                   & -    \\
    \hline
    \end{tabular}
\end{table*}

From TESS photometry, the ephemeris of the system has the following values:

\begin{eqnarray}
T_{c, e=0} &=& 2458764.31647 \pm 0.00018 \mathrm{BJD_{TDB}},\\ 
P_{e=0} &=& 1.2174928 \pm 1.7\times10^{-6}~ \mathrm{days}. 
\end{eqnarray}
The period value is in $1\,\sigma$ agreement to the most recent determination of \cite{Maciejewski2018}, but less precise due to our shorter time baseline.

\subsection{Updated stellar and companion parameters}
\label{sec:updated_params}

\begin{figure}[!ht]
    \centering
    \includegraphics[width=6.7cm,angle=90,trim=70 80 90 90,clip]{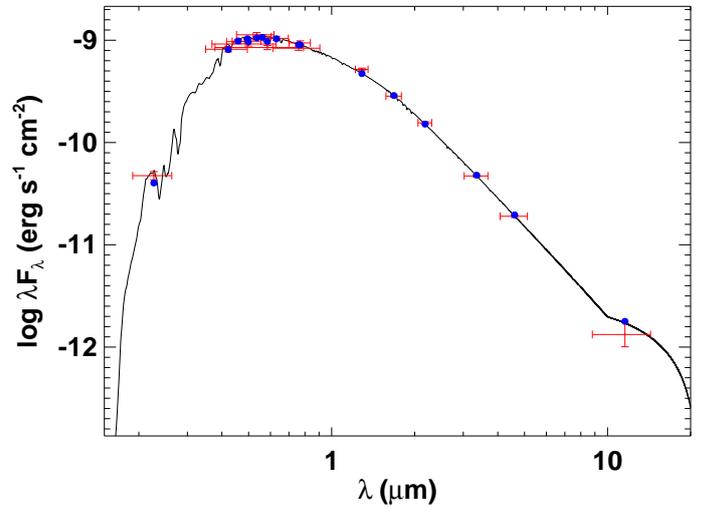}
\caption{Spectral energy distribution (SEDs) of KELT-1. Red symbols represent the observed photometric measurements, where the horizontal bars represent the effective width of the bandpass. Blue symbols are the model fluxes from the best-fit Kurucz atmosphere model (black). 
\label{fig:sed}}
\end{figure}

The stellar radius and stellar mass are varied as part of our modeling (see Section~\ref{sec:EV_DB}). As a consequence, in this work we have redetermined the stellar and brown dwarf companion radii and masses, taking advantage of the newly available parallax from {\it Gaia\/} DR2 \citep{GaiaCollaboration2018}, together with the available photometry from all-sky broadband catalogs. We used the semi-empirical approach of measuring the stellar spectral energy distribution (SED) described by \citet{Stassun:2017} and \citet{Stassun:2016}. We pulled the $B_T V_T$ magnitudes from {\it Tycho-2}, the $BVgri$ magnitudes from the APASS catalog, the $JHK_S$ magnitudes from {\it 2MASS}, the W1--W3 magnitudes from {\it WISE}, the $G G_{\rm BP} G_{\rm RP}$ magnitudes from {\it Gaia}, and the NUV magnitude from {\it GALEX}. Together, the available photometry spans the full stellar SED over the wavelength range 0.2--10~$\mu$m (see Fig.~\ref{fig:sed}). 

We performed a fit using Kurucz stellar atmosphere models, with the priors on effective temperature ($T_{\rm eff}$), surface gravity ($\log g$), and metallicity ([Fe/H]) from the spectroscopically derived values reported in the KELT-1b discovery paper \citep{Siverd2012}. The remaining free parameter is the extinction, $A_V$, which we limited to the maximum for the line of sight from the dust maps of \citet{Schlegel:1998}. The resulting fit (Fig.~\ref{fig:sed}) has a reduced $\chi^2$ of 1.6, and best-fit $A_V = 0.20 \pm 0.04$. Integrating the (unreddened) model SED gives the bolometric flux at Earth of $F_{\rm bol} = 1.616 \pm 0.038 \times 10^{-9}$~erg~s$^{-1}$~cm$^{-2}$. Taking the $F_{\rm bol}$ and $T_{\rm eff}$ together with the {\it Gaia\/} DR2 parallax, adjusted by $+0.08$~mas to account for the systematic offset reported by \citet{Stassun:2018}, we determined the stellar radius. From the empirical relations of \citet{Torres:2010} we estimated the stellar mass, which is consistent with that inferred from the spectroscopic $\log g$ together with the radius ($1.36 \pm 0.06$~$M_\odot$). Finally, the estimated mass together with the radius gives the mean stellar density. With the updated value for the stellar radius and the planet-to-star radii ratio derived fitting TESS photometry, we report here the KELT-1b radius in the TESS bandpass. All parameters derived here, and specified in Table~\ref{tab:StellarParams}, are in a 1-$\sigma$ agreement to the previous determination of \cite{Siverd2012} and yield a slight improvement in precision.

\begin{table}[ht!]
    \centering
    \caption{\label{tab:StellarParams} Updated stellar and brown dwarf companion parameters derived in this work.}
    \begin{tabular}{l c}
    \hline \hline
    Parameter                    &     Value \\
    \hline
    $R_S$ ($R_\odot$)        &    1.480 $\pm$ 0.028  \\
    $M_S$ ($M_\odot$)        &     1.34 $\pm$ 0.08   \\
    $\rho_S$ (~g~cm$^{-3}$)  &     0.58 $\pm$ 0.05   \\
    $R_P$ ($R_J$)                &    1.109 $\pm$ 0.021  \\
    \hline
    \end{tabular}
\end{table}

\section{Modeling}
\label{sec:modelling}

Our model consists of six components fit to the TESS data simultaneously. The components are presented in order of decreasing impact, which was determined by first fitting the different components to TESS data separately. To reduce the number of fitting parameters and to carry out a more realistic approach, if a given parameter was present in another model component, the pair was treated as equal. For each model component, we detail the model parameters, we highlight those that are being fit by bold-facing them in their respective equations, and we underline those that are being treated as equal by our MCMC algorithm. In all cases, the variable $t$ corresponds to time as measured and provided by TESS.

We carried out four fitting approaches, all of them considering zero eccentricity. First, we considered KELT-1b mass fixed to the value reported by \cite{Beatty2019} (henceforth, M1), and we used uniform priors in all the parameters except the orbital period and the mid-transit time of reference, which  always had Gaussian priors using the starting values and uncertainties reported in Table~\ref{tab:transit_parameters}. The difference between M1 and the second fitting approach, M2 is that M2 considers Gaussian priors on all the parameters. Then M3 and M4 are similar to M1 and M2, respectively, with the only difference that the brown dwarf mass is added as a fitting parameter. These four model approaches are intended to investigate whether the choice of priors has any impact on our results and to assess whether the TESS data would confidently allow for a proper mass determination through data fitting.

\subsection{Primary transit}
\label{sec:Pt}

We modeled KELT-1b's primary transits, PT(t), following \cite{MandelAgol2002}. The fitting parameters are, as specified in Section~\ref{sec:PT}, the semi-major axis, $a/R_S$, the inclination in degrees, $i$, the orbital period, $P$, the planet-to-star radius ratio, $R_P/R_S$, and the mid-transit time of reference, $T_c$. Thus, we have:\   

\begin{equation}
    \mathrm{PT(t) = PT(\underline{\bf a/R_S}, \underline{\bf i}, \underline{\bf P}, \underline{\bf R_P/R_S}, \underline{\bf T_c})}.
\end{equation}

\noindent The amplitude of the variation connected to the primary transit is its own depth, $\sim$5900 ppm. All parameters are fit and considered as equal by other model components.

\subsection{Ellipsoidal variation and Doppler beaming}
\label{sec:EV_DB}

The secondary eclipses presented by \cite{Beatty2019}, observed with Spitzer at 3.6 and 4.5~$\mu$m, are well placed nearby the maximum of planetary thermal emission. In constrast to this, after phase-folding the TESS KELT-1b data on its orbital period, we found the secondary eclipse to be placed at the minimum level of the phase curve variation (see Figure~\ref{fig:KELT1_final_model_data}, panel 2 from top to bottom). This is because the phase curve variability in TESS photometry is not dominated by the thermal emission of KELT-1b, as it is for the Spitzer light curves of \cite{Beatty2019}, but by the ellipsoidal variations in the host star caused by the gravitational pull of the brown dwarf companion. The amplitude of the effect depends on the properties of the stellar surface, the masses of the companion and the star, and the distance between the two bodies \citep[see e.g.,][]{Welsh2010,Mislis2012}. Even though we considered the eccentricity to be zero, the expression for ellipsoidal variation accounts for eccentric and inclined orbits \citep{Beatty2019}:
\begin{eqnarray}
\Delta F_{EV}(t) &=& A_{EV} \left( 1 - \cos\left(\frac{4\pi(t - \underline{\bf T_c})}{\underline{\bf P}}\right)\right),\,\\
A_{EV} &=& \beta\frac{\underline{\bf M_P}}{M_S}\underline{\bf \left(\frac{{\bf R_S}}{a}\right)}^3 \left(\frac{1 + e \cos(\nu)}{1 - e}\right)^3 \sin(\underline{\bf i})^3,\,\\
\beta &=& 0.12\frac{(15 + u)(1 + g)}{(3 - u)}.
\end{eqnarray}
\noindent In this equation, $M_P$ and $M_S$ correspond to the companion and stellar masses, respectively, $a/R_S$ is the relative distance between companion and star scaled to the stellar radius (semi-major axis for a circular orbit), and $\beta$ \citep{Morris1985} depends on the linear limb darkening coefficient $u$ and the gravity darkening coefficient $g$, which were taken from \cite{Claret2017}, interpolated to the stellar parameters of KELT-1. Specifically, for KELT-1b, $g = 0.2150$, and $u = 0.4337$. The mass of the star was adopted from this work (Section~\ref{sec:updated_params}). The fitting parameters are the mid-transit time, $T_c$, the orbital period, $P$, and the semi-major axis, $a/R_S$. The mass of the KELT-1b was fixed to the value of \cite{Beatty2019} in the modeling approaches M1 and M2, while in M3 and M4 we employed this mass estimation as prior. From literature values, the amplitude of the ellipsoidal variation is A$_{EV}\sim$400 ppm.

As KELT-1b and star orbit around their common barycenter, the star periodically moves towards and away from the observer with a frequency equal to the orbital period of the companion. This creates a variation of the host star brightness $\Delta F_{DB}$ due to Doppler beaming. Its expression \citep{Beatty2019} is given by:

\begin{eqnarray}
\Delta F_{DB}(t) &=& (3 - \alpha)\frac{K_{amp}(t)}{c},
\end{eqnarray}
\noindent where
\begin{eqnarray}
K_{amp}(t) &=& \frac{2\pi a_* \sin(\underline{\bf i})}{\underline{\bf P} \sqrt{1 - e^2}}\left[\cos(\omega + \nu_S(\underline{\bf T_c})) + e \cos(\omega)\right],\\
a_* &=& \left(\frac{\underline{\bf M_P}}{M_S}\right)\underline{\bf \left(\frac{a}{R_S}\right)}R_S.
\end{eqnarray}

\noindent Here, $\omega$ is the argument of the periastron, $\nu_S$ is the true anomaly of KELT-1, measured with respect to the mid-transit time, T$_0$, and R$_S$ is specified in Section~\ref{sec:updated_params}. The speed of light is represented by $c$ and $\alpha$ depends on the central wavelength of TESS bandpass and the effective temperature of the star \citep{Loeb2003}.  The fitting parameters are $a/R_S$, $T_c$, $P$, $M_P$ (only in M3 and M4), and $i$. The amplitude of the Doppler beaming is $K_{amp}\sim$90 ppm, based on the mass of KELT-1b derived in \cite{Beatty2019}. It differs from the ellipsoidal variation not only in amplitude, but also in frequency (one repetition per orbital period for the Doppler beaming, two repetitions per orbital period for the ellipsoidal variation). Like the ellipsoidal variations, Doppler beaming also strongly depends on the mass of the companion, hence we vary them together in this description.

\subsection{Reflected light, thermal emission and secondary eclipse}
\label{sec:SE_PC}

Following \cite{Cowan2008}, we model the reflection of starlight and the thermal emission of KELT-1b as a series of first order expansions in sines and cosines:

\begin{equation}
\mathrm{PPV(t)} = {\bf c_0} + {\bf c_1} \times \mathrm{\sin(2\pi t/\underline{\bf P})} + {\bf c_2} \times \mathrm{\cos(2\pi t/\underline{\bf P})}.
\end{equation}

\noindent PPV(t) stands for planetary phase variability. The fitting parameters of this model component are the vertical offset, c$_0$, the amplitudes of the sine and cosine, c$_1$ and c$_2$, respectively, and the orbital period, P. This linear combination of sines and cosines allow for a potential offset between the region of maximum brightness and the sub-stellar point. The amplitude of the phase curve is around $\sim$100 ppm. During secondary eclipse KELT-1b is hidden behind the star. As a consequence, the thermal emission and reflected light components should not show up during secondary eclipse. 

To model the secondary eclipse we make use of a step function \citep{Sackett1999}, in sync with the levels of PPV(t) right before and after secondary eclipse begins and ends, respectively, to allow for continuity. The shape of the step function is limited by the first to fourth contacts, which are in turn determined from the center of the eclipse, CE:

\begin{equation}
\mathrm{CE = \frac{\underline{\bf P}}{2}\times (1 + 4~e~ \cos(\omega))},\
\end{equation}

\noindent with the transit duration between first and fourth contacts:
\begin{equation}
\mathrm{T_{dur,1-4} =  \frac{\underline{\bf P}}{\pi} \sin^{-1}\left( \frac{\sqrt{ (1 + \underline{\bf R_P/R_s})^2 - (\underline{\bf a/R_S}~ \cos(\underline{\bf i}))^2}}{\underline{\bf a/R_S}}\right)},\
\end{equation}

\noindent and the transit duration between second and third contacts: 

\begin{equation}
\mathrm{T_{dur,2-3} =  \frac{\underline{\bf P}}{\pi} sin^{-1}\left( \frac{\sqrt{ (1 - \underline{\bf R_P/R_S})^2 - (\underline{\bf a/R_S}~ \cos(\underline{\bf i}))^2}}{\underline{\bf a/R_S}}\right)}.
\end{equation}

\noindent As the secondary eclipse is not necessarily symmetrically centered with the phase curve, the ingress and egress parts of the step function can have different levels. As a consequence, the eclipse depth ED is not measured as the difference of the out-of-eclipse flux to the in-eclipse flux, but as the difference between the point in the phase curve that corresponds to the center of the eclipse (as if the eclipse was not there), and the in-eclipse flux \citep[see e.g.,][their Figure 4]{Bell2019}. The secondary eclipse model depends on the following parameters:

\begin{equation}
\mathrm{SE(t) = SE(\underline{\bf a/R_S}, \underline{\bf i}, \underline{\bf P}, \underline{\bf R_P/R_S}, e, \omega, {\bf ED})}.
\end{equation}

\noindent Assuming pure thermal emission and thus no reflected light, using the brightness temperature determined by \cite{Beatty2019} at Spitzer bandpasses in the near IR, the expected secondary eclipse depth in the TESS bandpass is $\sim$300 ppm.

\subsection{Stellar activity}
\label{sec:activity}

After subtracting the primary transit, ellipsoidal variation and Doppler beaming models from the data using literature values, we hoped we would only be left with the reflected light, the thermal emission, and the secondary eclipse of KELT-1b. Visually inspecting the individual orbits, we noticed the shape of the phase curve and the eclipse depth were evidently changing from orbit to orbit. Before concluding that KELT-1b may be exhibiting weather patterns in its atmosphere \citep[see e.g.,][]{Armstrong2016}, we addressed other possible sources that could give rise to such variability. One of the sources investigated was stellar activity. As specified in \cite{Siverd2012}, the host is an F-type star on the main sequence with a rather shallow convective envelope. As a result, the star is expected to have a low level of activity or none at all. 

\begin{figure}[ht!]
    \centering
    \includegraphics[width=.5\textwidth]{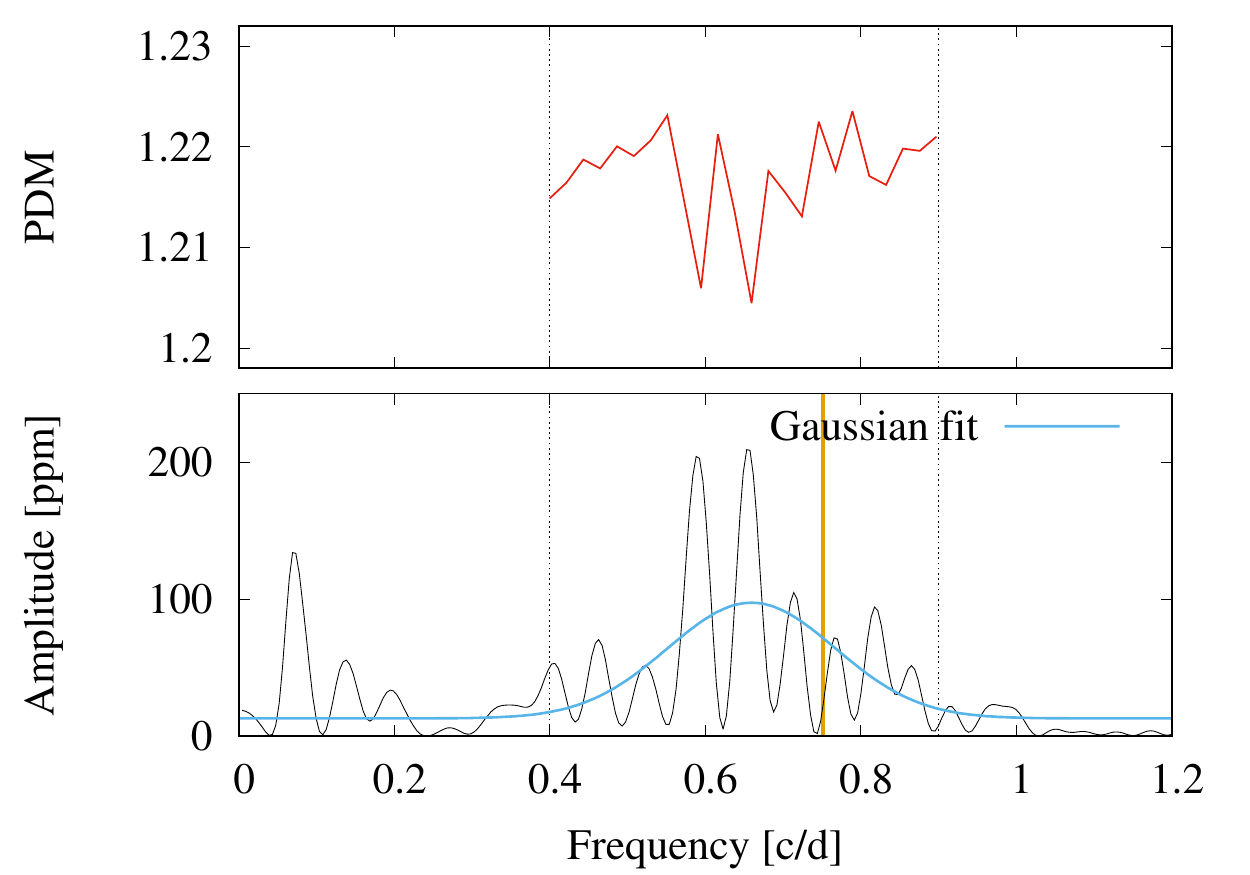}
    \caption{\label{fig:periodogram_PDM} {Periodicity of the residual light curve modulation. \it Top:} Phase dispersion minimization evaluated in the frequency range between 0.4 and 0.9 c/d. {\it Bottom:} Lomb-Scargle periodogram (amplitude in ppm versus frequency in cd$^{-1}$) for the residuals of KELT-1, showing only the frequency range where most of the power lies. Overplotted in blue is the fit of a Gaussian function over the power to estimate an error for $\nu_{min}$. The vertical orange line indicates the measured projected rotation velocity of the star as determined by \cite{Siverd2012}. For comparison, the frequency of the orbital motion of KELT-1b is 0.821~c/d.}
\end{figure}

In order to investigate the presence of spot modulation in the TESS data, we computed a Lomb-Scargle periodogram \cite{Lomb,Scargle,LombScargle} of TESS residuals, after the previously mentioned model components of Section~\ref{sec:Pt} to \ref{sec:SE_PC} were fit initially and removed. The resulting periodogram can be found in the bottom panel of Figure~\ref{fig:periodogram_PDM}. The figure shows unresolved peaks around 0.7 cycles per day (c/d), which could be due to rotational modulation induced by spots. Such a signal usually results in a peak of power around the rotational frequency of the star and will only show resolved peaks if the stellar spots are long-lived and do not migrate in latitude. As a consequence, we could not simply remove the peaks using  traditional frequency fitting. 

To determine the mean frequency of what appears to be rotational modulation of the host star, we used the phase dispersion minimization technique instead \citep[PDM,][]{Stellingwerf1978}. We divided the frequency space between 0.4 and 0.9 c/d, where most of the power lies, in steps of 1/2$\Delta$t, where $\Delta$t corresponds to the total time of TESS observations. Then, we converted these frequencies into periods (so-called trial periods), and with each one of these trial periods we phase-folded TESS residuals. We divided the orbital phase into 50 bins, and we computed the overall variance of all the samples per phase-folded dataset \citep[Equation 2 in][]{Stellingwerf1978}. We compared this value to the variance of the residuals. If the trial period minimizes the dispersion of the phase-folded data, then the PDM should be minimized. The top panel of Figure~\ref{fig:periodogram_PDM} shows the overall variances for the different trial periods, converted back to frequencies. The minimum corresponds to \mbox{$\nu_\mathrm{min,spots}$ = 0.659 cd$^{-1}$}, slightly rightward of the highest peak. This results was consistent even after changing the frequency range where the PDM was evaluated, and the number of bins into which we divided the phase-folded data. To estimate an uncertainty for this value we carried out a fit of the spectral power with a Gaussian function. To derive the best-fit values, we carried out a least-squared minimization between data and model. For the Gaussian function, we considered the mean value as fixed, and equal to $\nu_\mathrm{min,spots}$. The fitting parameters are the amplitude, the offset, and the standard deviation. From the latter, we computed the error of $\nu_\mathrm{min,spots}$ to be equal to 0.107 cd$^{-1}$. The fit is shown in Figure~\ref{fig:periodogram_PDM} with a blue continuous line. If this periodicity is interpreted as the stellar rotation period with $P=1.52\pm0.29$~d, we find an agreement within 1~$\sigma$ to the stellar period derived from the measured projected rotation velocity of the star of $P=1.33\pm0.06$~d \citep{Siverd2012}.

To represent the observed stellar activity (SA) of KELT-1 we modeled it as follows:

\begin{equation}
    SA(t) = {\bf A_{spots}} \times \sin[2\pi(t/\nu_{min,spots} + {\bf \boldsymbol\phi_{spots}})],\,
\end{equation}

\noindent where A$_\mathrm{spots}$ corresponds to the amplitude of the photometric variability due to stellar spots, $\nu_\mathrm{min,spots}$ corresponds to the frequency minimizing the PDM, and $\phi_\mathrm{spots}$ is the phase, in units of 2$\pi$. As the periodogram shows, the amplitude of this effect is A$\sim$210 ppm. The fitting parameters are the amplitude and the phase. We do not fit for the frequency, as we are limited by the total amount of time of TESS on the target.

\subsection{Combined model and fitting approach}

Our combined model is the addition of the six components previously described, PT(t) + $\Delta$F$_{EV}$(t) + $\Delta$F$_{DB}$(t) + SA(t) + PPV(t) + SE(t). For M1 and M2, the total number of fitting parameters are 11, while for M3 and M4 they are 12, when the mass of the brown dwarf is included. We note that the vertical offset is arbitrarily included in the PPV(t). As our combined model is a simple linear combination of the six components, it is irrelevant where the offset is. This combined model was fit to the unbinned TESS photometry. For each one of the model approaches, we determined the BIC to determine which fitting approach was the best at reproducing the data.

As in the primary transit fitting approach, to derive the best-fit values for the parameters, we made use of the MCMC. In each one of the previously mentioned cases we iterated \mbox{10\ 000} times, with a burn-in of the first 2000 samples. This burn-in was determined from a posterior visual inspection of the chains, where we also investigated their convergence. As usual, the best-fit values for the parameters, along with their corresponding 1-$\sigma$ uncertainties, are derived from the mean and standard deviation of the posterior distributions.
 
\section{Results}
\label{sec:results}

\subsection{Parameters derived for KELT-1b}
\label{sec:params_KELT1_derived}

\begin{figure*}[ht!]
    \centering
    \includegraphics[width=.85\textwidth]{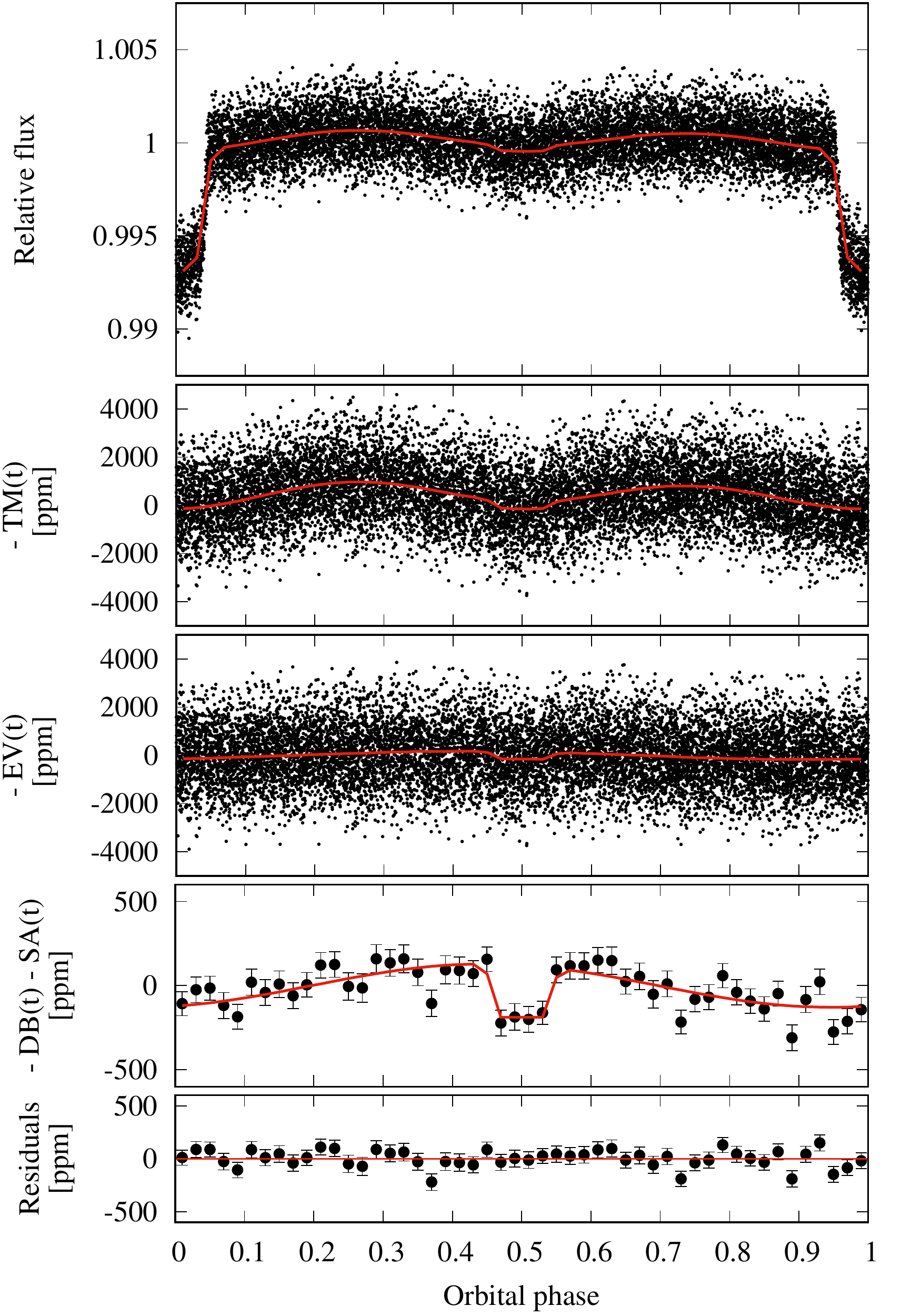}
    \caption{\label{fig:KELT1_final_model_data} Data (black points) versus best-fit model (red continuous line) as a function of the orbital phase of KELT-1b. From top to bottom: Full data set; same data, but with the best-fit primary transit, PT(t), removed; ellipsoidal variation, EV(t), removed; Doppler beaming and stellar activity, DB(t) and SA(t), respectively, removed, at which point the secondary eclipse and the phase curve are visible; residuals, once the secondary eclipse and the phase curve are removed. The last two panels are shown at a 0.02 orbital phase binning. The last four panels are given in ppm.}
\end{figure*}

Table~\ref{tab:results_MCMC} shows the parameters obtained from our combined fit, along with those determined from these values. The corresponding best-fit values for the derived parameters, along with their uncertainties, were computed in the usual way from their posterior distributions. TESS phase-folded data, along with our best-fit combined model, can be seen in Fig.~\ref{fig:KELT1_final_model_data} for M2, which is the model choice with the lowest BIC value. The figure shows TESS photometry of KELT-1b in black points, phase-folded with the best-fit orbital period. Primary transits are close to phases \mbox{$\phi$ = 0 and 1}, while the secondary eclipses lie around \mbox{$\phi$ = 0.5}. The first three panels show TESS unbinned data, while the last two are binned each \mbox{$\Delta\phi$ = 0.02} (equivalently, $\Delta$t$\sim$30 minutes). At this cadence, the photometric precision is 80 ppm. 

From the coefficients connected to the PPV(t) component we determined the amplitude of the phase curve: 

\begin{equation}
    \mathrm{A(c_1, c_2) = \sqrt{c_1^2 + c_2^2}}.
\end{equation}

\noindent In addition, we computed the offset between the region of maximum brightness and the substellar point:

\begin{equation}
   \phi_\mathrm{off} = \tan^{-1}(c_1/c_2).
\end{equation}

\noindent The derived values for the different fitting approaches can be seen in Table~\ref{tab:results_MCMC}.

\subsection{Physical parameters derived from these observations}
\label{sec:params_W33_derived_2}

We determined the equilibrium temperature at the sub-stellar point on KELT-1b (i.e., the irradiation temperature),

\begin{equation}
    T_0 = T_{\rm eff} \sqrt{R_S/a}\,
,\end{equation}

\noindent where $T_\mathrm{eff}$ is the effective temperature of the star and $a/R_S$ is the semi-major axis scaled by the stellar radius obtained from the primary transit fitting component of our joint model \citep{MandelAgol2002}. From this parameter, we can derive the effective temperature of the day side in the no-albedo, no-circulation limit, $T_{\epsilon = 0} = (2/3)^{1/4}T_0$ \citep{Burrows2008b,Cowan2011}, the maximum possible day-side temperature of KELT-1b.

To compute the brightness temperature, $T_b(\lambda)$, on the day and the night side of KELT-1b, we follow  \cite{Cowan2011},
\begin{equation}
T_b(\lambda) = \frac{hc}{\lambda k}\left[\log\left(1 + \frac{\exp\big(hc/\lambda k T_{*b}(\lambda)\big)-1}{\psi(\lambda)}\right)\right]^{-1}.
\label{eq:C&A2011}
\end{equation}
Here, $T_{*b}(\lambda)$ is the star's brightness temperature in the TESS bandpass, and $h$, $c$ and $k$ are the Planck constant, the speed of light, and the Boltzmann constant, respectively. For KELT-1, the brightness temperature is $T_{*b} = 6456$~K. In the equation, $\psi(\lambda)$ corresponds to the ratio of the companion's day- or night-side intensity to the stellar intensity at that wavelength. In particular, for the day-side temperature we made use of the secondary eclipse depth divided by the primary transit depth, as this is a direct measure of the ratio of the companion's day-side intensity to the stellar intensity, 

\begin{equation}
    \psi(\lambda)_{day} = ED/(R_P/R_S)^2.
\end{equation}

\noindent To compute the night-side temperature we used the difference between the secondary eclipse depth and the phase variation amplitude, combined with the offset between the peak of the phase curve and secondary eclipse: 

\begin{equation}
    \psi(\lambda)_{night} = \frac{ED - 2A \cos\phi_\mathrm{off}}{(R_P/R_S)^2}.
\end{equation}

\begin{figure*}[ht!]
    \centering
    \includegraphics[width=.95\textwidth]{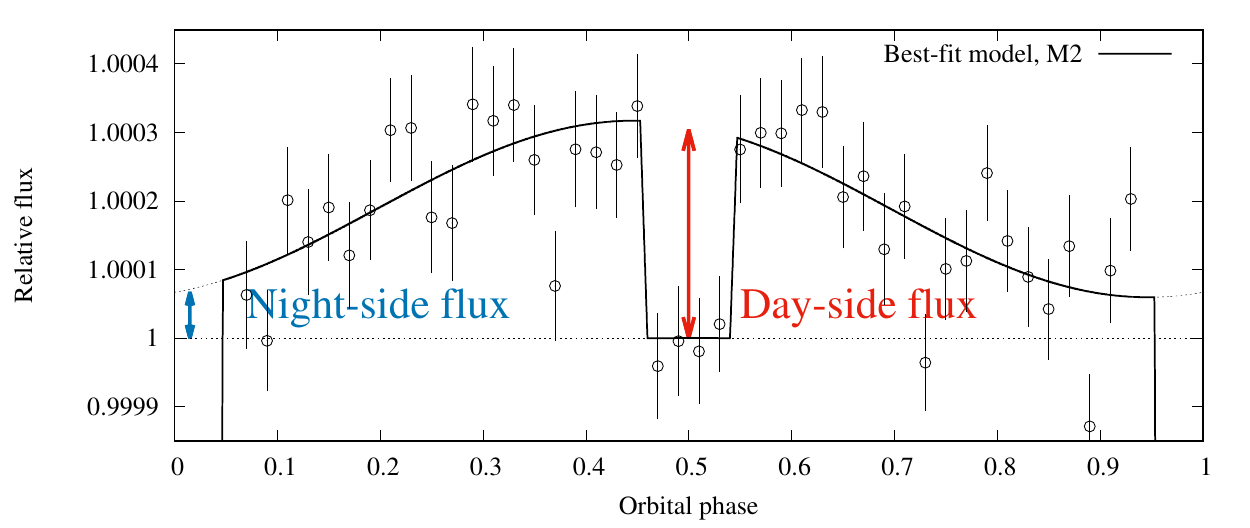}
    \caption{\label{fig:KELT1_final_model_data_zoom} Relative flux of TESS data binned as specified before, showing only the phase curve and secondary eclipse components. In red and blue bars, we indicate the day- and night-side fluxes.}
\end{figure*}

\noindent We then computed the brightness temperatures of the day and the night side of KELT-1b by substituting these two values into Equation~\ref{eq:C&A2011}.
The lowest point of the phase curve in Figure~\ref{fig:KELT1_final_model_data_zoom} shows a slightly higher flux level than the bottom of the secondary eclipse, which suggests a night-side emission from KELT-1b in the TESS bandpass. However, this estimated amount of night-side flux is significant to less than 2~$\sigma$, therefore, the derived night-side temperature $T_{\rm night}$ of about 1400~K is not well constrained either.

We obtained uncertainties on all these parameters using the posterior probability distributions for the values $c_1$, $c_2$, $ED$ and $a/R_S$. In particular, for each one of the 8000 MCMC iterations, we computed each one of the previously specified equations. In this way, the values reported in Table~\ref{tab:results_MCMC} have their best-fit values and uncertainties determined from their mean and standard deviations. 

Table~\ref{tab:results_MCMC} summarizes the parameters derived in this work and compares them to those computed by \cite{Beatty2019}. To begin with, the offset between the peak of the phase curve and secondary eclipse is consistent within 1-$\sigma$ to the value reported by \cite{Beatty2019}, regardless of the fitting approach we used. 
This suggests that the TESS phase curve is sensitive to thermal emission of KELT-1b, as is Spitzer. However, the day-side brightness temperature inferred from TESS data is higher than this reported by \cite{Beatty2019}. The difference may be attributed to stronger molecular absorption in the Spitzer IRAC bands compared to that in the TESS band, as discussed in Section~\ref{sec:models}.

The day-side brightness temperature in the TESS bandpass is marginally larger than $T_{\epsilon=0}$\footnote{The disk-integrated day-side effective temperature of KELT-1b cannot significantly exceed this limit even in the presence of internal heat, since $T_{\rm max} = (T_{\epsilon=0}^4 + T_{\rm int}^4)^{1/4} \approx T_{\epsilon=0}$.}, even though it still consistent at the 1-$\sigma$ level. This might suggest that either: a) the two measurements are indeed consistent, and we can derive reliable Bond albedo and heat redistribution efficiency for KELT-1b (Section~\ref{sec:Spitzer_TESS}); b) reflected light contributes to the large TESS eclipse depth (Section~\ref{sec:RL}); or c) an emission feature in the TESS bandpass increases the brightness temperature above the effective temperature. Here, the TESS data could be probing the deeper hotter layers of the atmosphere compared to the Spitzer bands, which could have strong opacity and, hence, probe cooler upper regions of the atmosphere (Section~\ref{sec:models}). We discuss these options in the following sections.

\begin{table*}[ht!]  
    \centering
    \caption{\label{tab:results_MCMC} Best-fit values and 1-$\sigma$ uncertainties derived from our four approaches. From top to bottom, we show the eclipse depth, ED, the coefficients accompanying the phase curve model (c$_0$, c$_1$, and c$_2$), and those with the stellar activity model, A$_{spots}$ and $\phi_{spots}$. In addition, the amplitude of the phase curve, A, the offset between the region of maximum brightness and the sub-stellar point, $\phi_\mathrm{off}$, the irradiation temperature, T$_o$, the temperature of the day side in the no-albedo, no-circulation limit, T($\epsilon$ = 0), and the day- and night-side temperatures, T$_\mathrm{day}$ and T$_\mathrm{night}$, respectively. The last row shows the BIC, and the last column specifies relevant values obtained by \cite{Beatty2019}. Two values separated by a slash correspond to the Spitzer bandpass of 3.6 and 4.5 $\mu$m, respectively.}
    \begin{tabular}{l c c c c c}
    \hline\hline
    Parameter             & M1                 &  M2                &  M3               &  M4                &  \cite{Beatty2019}    \\
                          &                    &                    &                   &                    &  (Spitzer, 3.6/4.5 $\mu$m) \\
    \hline
    $ED$                    & 308 $\pm$ 69       & 304 $\pm$ 75       & 320 $\pm$ 69      & 327 $\pm$ 72       &  -  \\
    $c_0$ (ppm)           & -284 $\pm$ 29      & -298 $\pm$ 32      & -279 $\pm$ 38     & -286 $\pm$ 27      &  -  \\
    $c_1$ (ppm)           & -46.6 $\pm$ 22.3   & -42.4 $\pm$ 20.8   & -57.4 $\pm$ 25.4  & -44.6 $\pm$ 22.3   &  -  \\
    $c_2$ (ppm)           & 122.4 $\pm$ 29.0   & 121.6 $\pm$ 27.9   & 119.8 $\pm$ 28.9  & 129.0 $\pm$ 28.1   &  -  \\
    $A_\mathrm{spots}$ (ppm)& 195.3 $\pm$ 23.8 & 195.3 $\pm$ 23.5   & 194.2 $\pm$ 23.6  & 194.5 $\pm$ 23.1   &  -  \\
    $\phi_\mathrm{spots}$ (2$\pi$)& 0.04 $\pm$ 0.01 & 0.05 $\pm$ 0.02 & 0.05 $\pm$ 0.02 & 0.05 $\pm$ 0.02    &  -  \\
    \hline\hline
    $A$                     & 131.1 $\pm$ 28.4   & 128.7 $\pm$ 27.2   & 132.9 $\pm$ 28.2  & 136.5 $\pm$ 27.5   & 959 $\pm$ 41     \\
    $\phi_\mathrm{off}$ ($^{\circ}$) & -20.8 $\pm$ 9.9 & -19.2 $\pm$ 9.6 & -25.6 $\pm$ 11.3& -19.1 $\pm$ 9.6 & -28.6 $\pm$ 3.8  \\
    $M_{\rm P}$ (M$_J$)  & 27.23              &  27.23             & 22.1 $\pm$ 3.4    & 27.2 $\pm$ 0.5     & 27.23 $\pm$ 0.50 \\
    \hline
    $T_0$ (K)             & 3424 $\pm$ 51      & 3407 $\pm$ 69      & 3528 $\pm$ 157    & 3395 $\pm$ 54      & 3391 $\pm$ 31   \\
    $T_{\epsilon=0}$ (K) & 3094 $\pm$ 25      & 3079 $\pm$ 28      & 3188 $\pm$ 83     & 3067 $\pm$ 23      & 3064 $\pm$ 28    \\
    $T_{\rm day}$ (K)         & 3209 $\pm$ 141     & 3201 $\pm$ 147     & 3213 $\pm$ 136    & 3247 $\pm$ 134     & 2988 $\pm$ 60/2902 $\pm$ 74 \\
    \hline
    BIC                   & 5215.7             &  5215.4            &  5227.1              &   5225.2           & -    \\           
    \hline
    \end{tabular}
\end{table*}

\section{Discussion}\label{sec:context}

\subsection{Bond albedo and heat redistribution efficiency through a self-consistent analysis combining Spitzer \& TESS}
\label{sec:Spitzer_TESS}

To derive the Bond albedo and the heat redistribution efficiency for KELT-1b, we assume that the flux of KELT-1b measured by TESS is solely composed of thermal emission. Furthermore, we make the simplifying assumption that the atmosphere of KELT-1b behaves entirely as a blackbody. In other words, we can infer a Bond albedo that neglects the fact that opacities might be caused by molecular absorption. Potential deviations from this blackbody assumption are considered in Section~\ref{sec:models}. We estimate the day-side temperature using the error-weighted mean of the Spitzer and TESS brightness temperatures, that is, by combining the \cite{Beatty2019} values specified in the last column of Table~\ref{tab:results_MCMC} and our values in the TESS bandpass for M2, while for the night-side effective temperature we only average the two Spitzer measurements because of the unconstrained TESS value. The resulting day and night effective temperatures are listed in Table~\ref{tab:results_averages}. The averaged day-side temperature is close to the value of \cite{Beatty2019}, since their more precise Spitzer results dominate the error-weighted mean. It is worth mentioning that T$_\mathrm{day}$ is below $T_{\epsilon=0}$.

\begin{table}[htb!]  
    \centering
    \caption{\label{tab:results_averages} Averaged day- and night-side temperatures, along with the Bond albedo, A$_B$, and the heat redistribution efficiency, $\epsilon$, determined from these values under the simplifying assumption that KELT-1b radiates as an isothermal blackbody.}
    \begin{tabular}{l c}
    \hline\hline
    Parameter              & Value         \\
    \hline
    T$_\mathrm{day}$ (K)   & 2996 $\pm$ 44 \\
    T$_\mathrm{night}$ (K) & 1128 $\pm$ 120 \\
    T$_0$ (K)              & 3405 $\pm$ 47 \\
    A$_B$ (\%)             & 7.1 $\pm$ 7.7 \\
    $\epsilon$             & 0.052 $\pm$ 0.006 \\
    \hline
    \end{tabular}
\end{table}

We then follow \cite{Cowan2011} to determine the Bond albedo: 
\begin{equation}
    A_B = 1 - \frac{5 T_{\rm night}^4 + 3 T_{\rm day}^4}{2T_0^4}\,,
    \label{eq:Ab}
\end{equation}

\noindent and the heat redistribution efficiency,
\begin{equation}
    \epsilon = \frac{8}{5 + 3 (T_{\rm day}/T_{\rm night})^4} .
    \label{eq:eps}
\end{equation}

The resulting parameters are also listed in Table~\ref{tab:results_averages}. 

\subsection{Reflected light to explain the deep TESS eclipse}
\label{sec:RL}
Since the amount of thermally emitted light by KELT-1b is reduced at the optical TESS bandpass compared to the Spitzer bandpasses, a potential contribution of reflected light to the secondary eclipse depth would have a considerable impact on the calculation of the brightness temperature. Therefore, it is tempting to attribute the high TESS day-side brightness temperature to reflected light, which would not have significantly affected the Spitzer measurements. To investigate this possibility, we then model the TESS eclipse depth following \cite{Alonso2018}:

\begin{equation}
    \frac{F_P}{F_S} = A_g \left(\frac{R_P}{a}\right)^2 \phi(\alpha) + \frac{B(\lambda, T_{d,p})}{B(\lambda,T_s)} \left(\frac{R_P}{R_S}\right)^2 ,
\end{equation}

\noindent where the first term of the right-hand side of the equation corresponds to reflected light and the second term is thermal emission. For the reflected light term, $A_g$ is the geometric albedo, $R_P$ and $R_S$ are the planet and stellar radius, respectively, $a$ is the semi-major axis, and $\phi(\alpha)$ is the phase function that can be approximated to 1 during secondary eclipse. For the thermal emission term, $B(\lambda, T_{d,p})$ and $B(\lambda,T_s)$ correspond to the blackbody emission of the companion and star, respectively, at brightness temperatures of $T_{d,p}$ and $T_s$. If we assume the brightness temperature in the TESS bandpass to be the same as measured in the Spitzer bandpasses, \mbox{$T_{d,p} = 2988 \pm 60$ K} \citep{Beatty2019}, then using our derived values for the TESS eclipse depth, the planet-to-star radii ratio, and the semi-major axis scaled to the stellar radius, we estimate a geometric albedo of 25 $\pm$ 11\%. Hot Jupiter exoplanets have been observed with a range of geometric albedos from low values close to zero \citep{Rowe2008,Esteves2015,Mallonn2019} to values of up to 0.3 and more \citep{Demory2011,Keating2017,Wong2020b}. A large geometric albedo at optical wavelengths has for example been observed for Kepler-7b with $A_g = 32 \pm 3$\% \citep{Demory2011}. The authors interpreted this large geometric albedo as caused by Rayleigh scattering. 

Our day and night temperatures satisfy $T_{\rm day}^4 + T_{\rm night}^4 < \frac{2}{3}T_0^4$. This means that KELT-1b has a non-zero Bond albedo, if the day-side and night-side atmospheres are treated as blackbodies \citep{Cowan2011}. In order to evaluate whether the Bond and geometric albedos are self-consistent, we estimate the fraction of incident stellar light in the TESS bandpass: we convolved and afterwards integrated the TESS transmission response with PHOENIX stellar intensity models for $T_{\rm eff} = 6500$~K, $\log(g) = 4.5$ and [Fe/H] = 0, matching the values of KELT-1 within uncertainties. The fraction of stellar flux in the TESS bandpass is $F_\mathrm{TESS}$ = 31\%. The minimum Bond albedo, $A_{B,min}$ should be $A_g \times F_\mathrm{TESS} = 7.6$\% when considering the geometric albedo specified in Section~\ref{sec:RL}. This value is compatible with the Bond albedo determined in Section~\ref{sec:Spitzer_TESS} (Table~\ref{tab:results_averages}), but requires the geometric albedo to be nearly zero outside of the TESS bandpass.

Even though this analysis of the Bond and geometric albedo of the day side is self-consistent, the assumption of pure blackbody radiation with the day-side brightness temperature being the same at TESS and Spitzer bandpasses might be overly simplified. The depth of the secondary eclipse at optical wavelength could also be explained by the TESS band probing hotter layers of the atmosphere compared to those probed by Spitzer, due to a non-isothermal temperature profile and wavelength-dependent opacity. This last scenario is first motivated (Section~\ref{sec:Teff_estimate}) and investigated (Section~\ref{sec:models}) in the following subsections.

\subsection{Initial estimation of the effective temperature of KELT-1b}
\label{sec:Teff_estimate}

To determine the effective temperature of KELT-1b, we assume at this point that KELT-1b's emission can be approximated by a blackbody. Under this assumption, we fit all the literature values reported by \cite{Beatty2014,Beatty2017,Beatty2019} and \cite{Croll2015} to our new measurement to synthetic eclipse depths. These values were determined by first computing the flux ratio between KELT-1b, represented by a blackbody function of different temperatures, and PHOENIX intensities, following the stellar parameters specified before, to represent KELT-1. This flux ratio was then integrated within all the transmission functions for which literature data were obtained. For the blackbody we created a grid of temperatures between 2000 and 4000 K, with a step of 1 K. For each one of these temperatures we computed the previously mentioned ratio, after which we determined the $\chi^2$ between the observations and our model. From a $\chi^2$ minimization, we obtained \mbox{T$_\mathrm{eff}$ = 3010 $\pm$ 78 K}. The error on the temperature was computed considering $\Delta\chi^2 = 1$.

\begin{figure}[ht!]
    \centering
    \includegraphics[width=.5\textwidth]{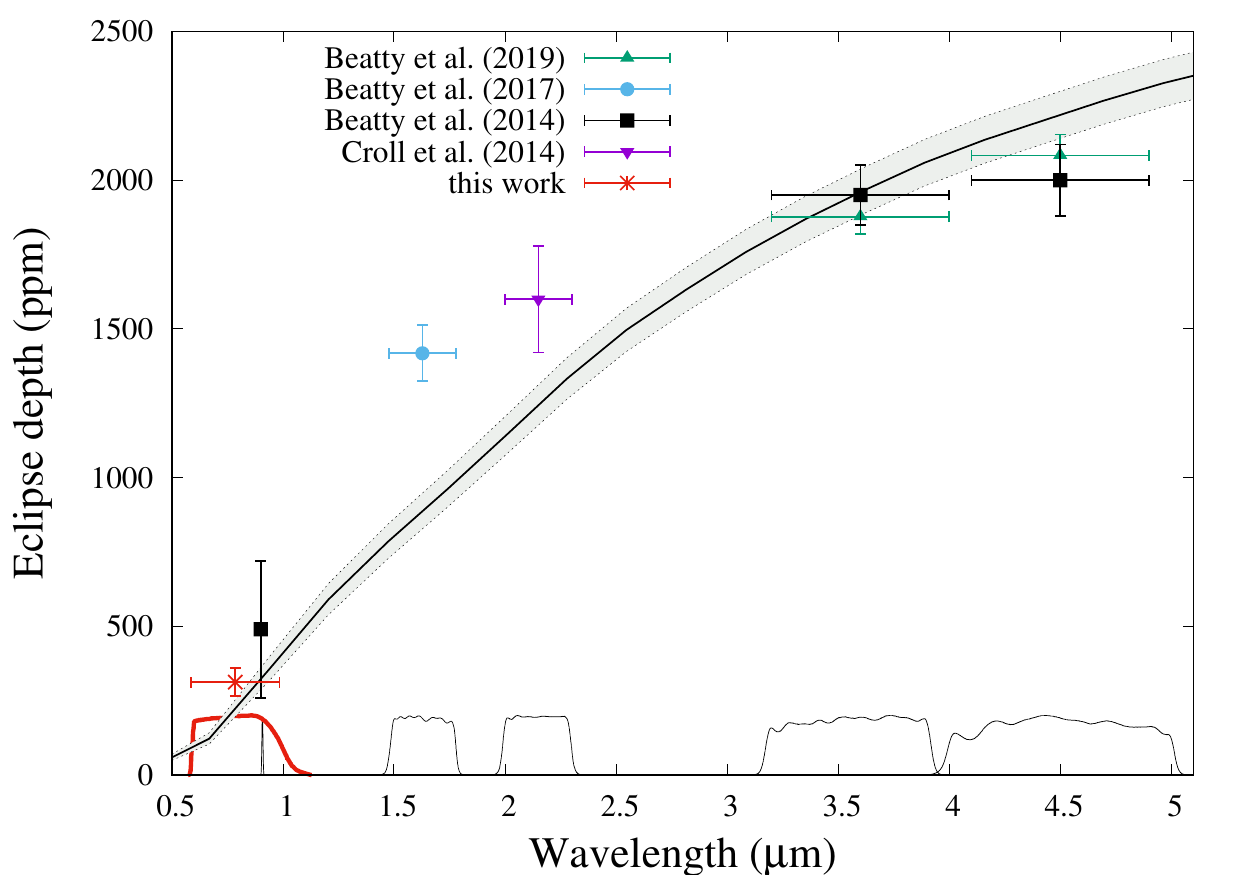}
    \caption{\label{fig:Teff_from_ED} Eclipse depth, in ppm, as a function of wavelength. Literature values are taken from \cite{Croll2015,Beatty2014,Beatty2017,Beatty2019}. Our new data point is plotted in red. The gray area shows a 1-$\sigma$ contour of the equilibrium temperature of KELT-1b, determined from all data. Transmission responses are plotted in black continuous lines, with the exception of TESS, highlighted in red.}
\end{figure}

At first glance, the figure reveals that both the \cite{Beatty2017} and \cite{Croll2015} eclipse depth measurements are significantly higher compared to the derived effective temperature. Both measurements, along with the z-band measurement reported by \cite{Beatty2014}, share the same observing technique: they were derived from incomplete orbital phase coverage, solely from secondary eclipse observations. As the mass of KELT-1b is particularly high, the ellipsoidal variation of the star creates a convex secondary eclipse. If the ellipsoidal variation is not removed, then the measured secondary eclipse, due to this convexity, will be deeper \citep[see Section 4.4 of][]{Bell2019}. To quantify how much deeper, we carried out the following exercise, focusing on the H-band data reported by \cite{Beatty2017}. First, we created synthetic secondary eclipse data. We used the parameters, cadence, and noise amplitude from  \cite{Beatty2017}. However, for this exercise, the noise was purely white. The synthetic data include the secondary eclipse, some thermal emission, the ellipsoidal variation, and the Doppler beaming. Once the synthetic time stamps, data and errors were generated, we fit them. As a model, in this case, we used a second-order time-dependent polynomial and a step function with a flat bottom as the secondary eclipse. The detrending should take into account every effect introduced in generating the synthetic data. We fit the data and model using least square minimization. The fitting parameters featured the detrending coefficients and the secondary eclipse depth. We repeated the same process 1\ 000 times. Within these repetitions, we simply counted how many times was the derived secondary eclipse depth larger than the one we used to create the synthetic data, equal to the value reported in \cite{Beatty2017} for the H band. Regardless the number of repetitions we carried out, we always obtained the same results. About 80\% of the time, the fitted secondary eclipse depth was larger than the input one. As a result, the fit is clearly opting for a deeper eclipse to compensate for the convexity. In addition, the percentage at which the secondary eclipse depth was overestimated is around 10\%. In other words, by not taking into account the gravitational pull of KELT-1b induced over the star, and the effect on the secondary eclipse light curve, the secondary eclipse depth is enlarged by about 10\%. 

When determining the effective temperature of KELT-1b without considering the H- and K- band measurements, the effective temperature decreases to \mbox{$T_\mathrm{eff} = 2946 \pm 71$~K}, and $\chi^2_{\rm red}$ decreases from 6.8 to 1.6. As we believe the eclipse depth measurements obtained from incomplete phase curve analysis are overestimated, in order to further carry out a more detailed analysis of the atmospheric properties of KELT-1b, we go on to focus on the TESS and Spitzer measurements, both of which benefit from full-orbit monitoring.

\subsection{Self-consistent atmospheric models}
\label{sec:models}

\begin{figure}
    \centering
    \includegraphics[width=0.5\textwidth]{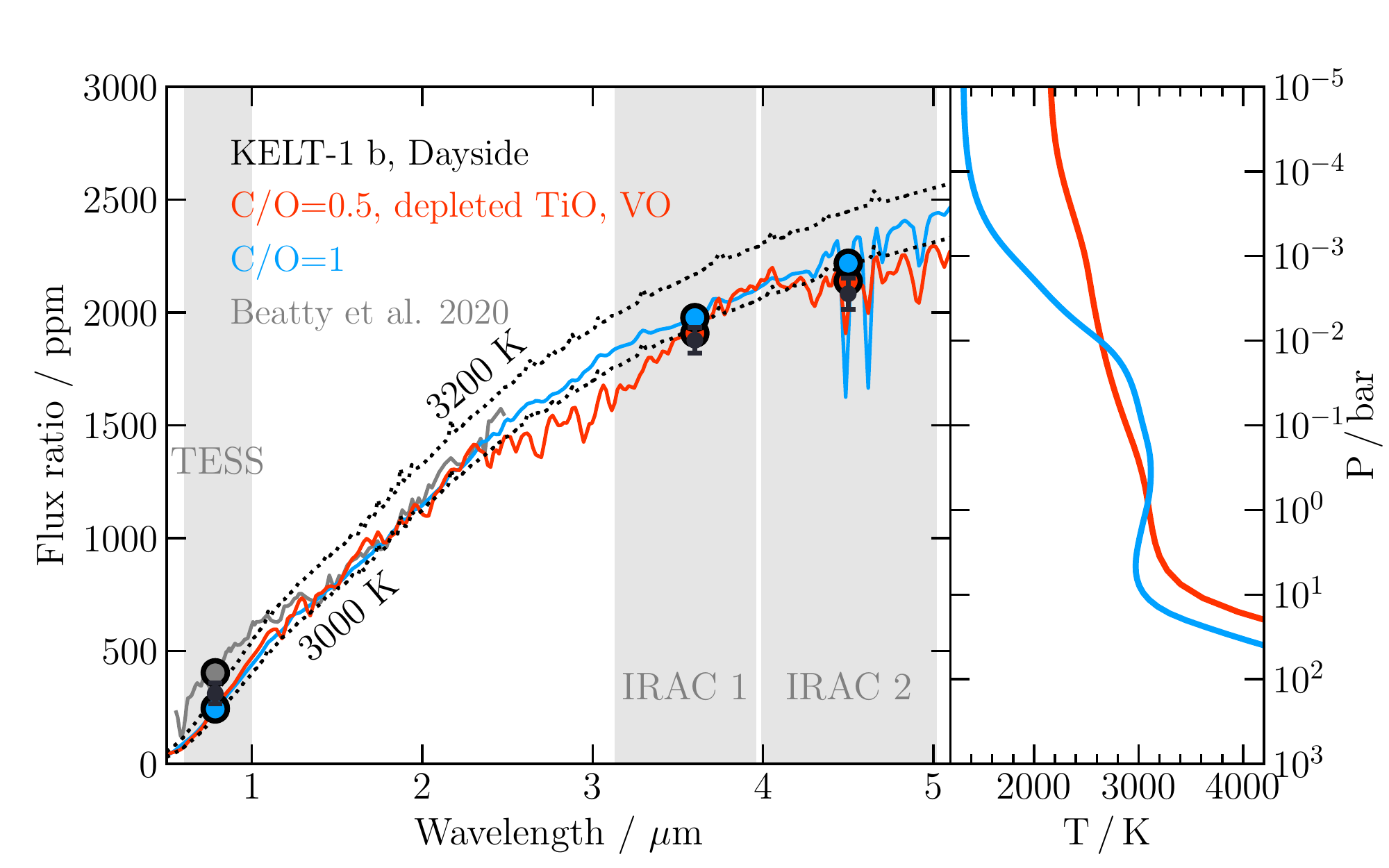}
    \includegraphics[width=0.5\textwidth]{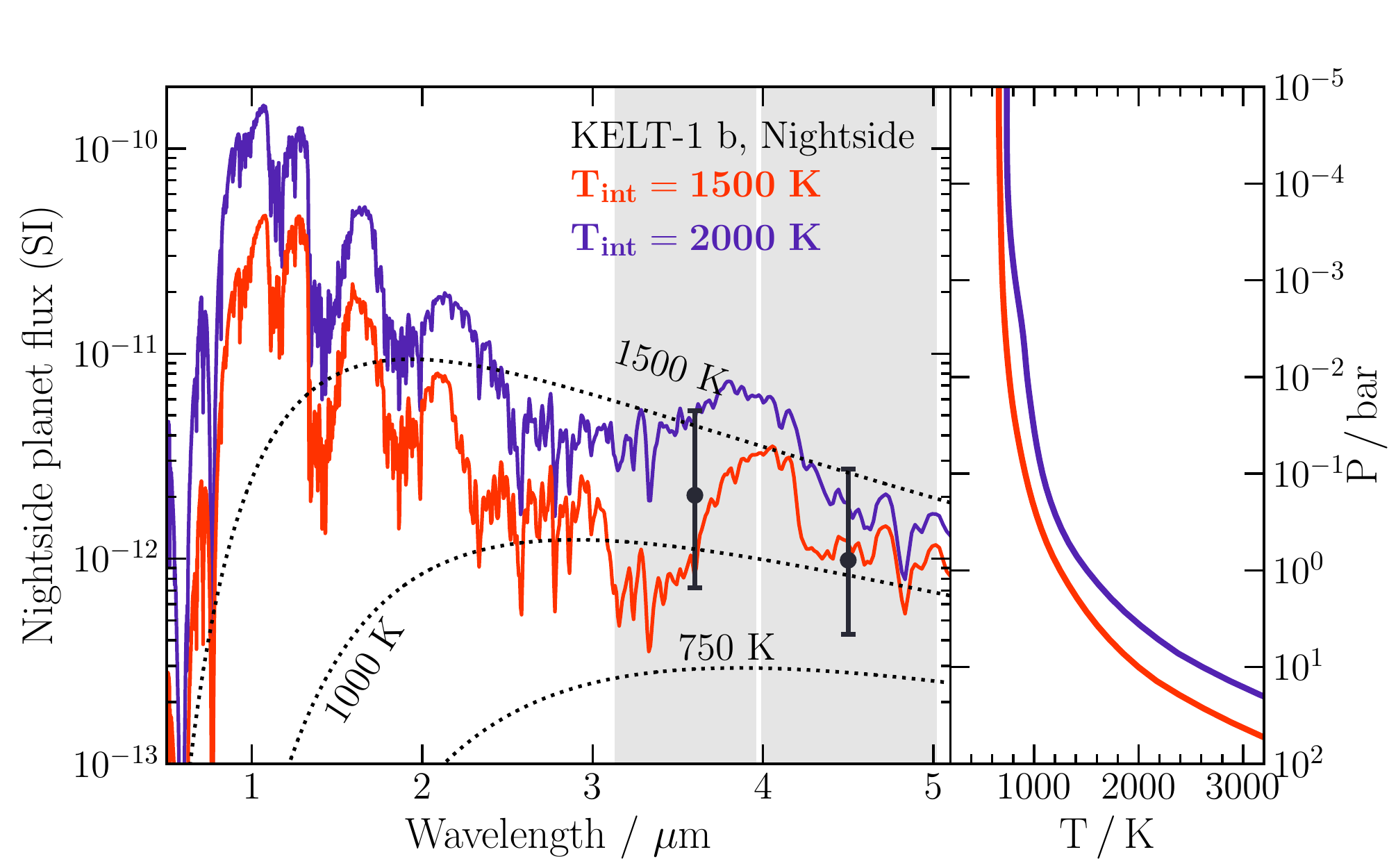}
    \caption{ {Self-consistent modeling of the atmosphere of KELT-1b. \it Top:} Model day-side spectra (left) and pressure-temperature ($P$-$T$) profiles (right), assuming solar metallicity and an internal temperature of 1500~K. The red spectrum and $P$-$T$ profile correspond to a C/O=0.5 and depleted TiO/VO (e.g., due to gravitational settling), while the blue model corresponds to a C/O=1. The black dashed lines correspond to blackbody spectra at 3000 and 3200~K. The cloudy, T$_{\rm int}$=1000~K model from \citet{Beatty2020} is shown in grey for reference. {\it Bottom:} Model night-side spectra (left) and pressure-temperature profiles (right) for KELT-1~b assuming internal temperatures of 1500~K (red lines) and 2000~K (purple lines). Dashed black lines correspond to blackbody spectra at 750, 1000 and 1500~K. Flux is shown in SI units (W sr$^{-1}$m$^{-3}$). In both panels, data are shown by black points and error bars. Coloured circles show the binned model spectra. Vertical gray bands show the TESS and Spitzer bandpasses.
    \label{fig:atm_models}}
\end{figure}

In this section, we self-consistently model the day- and night-side spectra of KELT-1~b using the \textsc{Genesis} atmospheric model \citep{Gandhi2017}. \textsc{Genesis} calculates full, line-by-line radiative transfer under radiative-convective, local thermodynamic, hydrostatic, and thermochemical equilibrium. The model self-consistently calculates the atmospheric temperature profile, thermal emission spectrum, and chemical profiles given the incident irradiation, internal flux, and elemental abundances (via metallicity and C/O ratio). Equilibrium chemical abundances are calculated using the \textsc{hsc chemistry} (version 8) software package, as in \citet{Piette2020}. \textsc{hsc chemistry} uses the \textsc{gibbs} solver \citep{White1958} to calculate chemical abundances by minimising the total Gibbs free energy of the system and includes effects due to thermal dissociation or ionisation, which are relevant with regard to to the highly-irradiated day side of KELT-1~b. 

While the equilibrium chemistry calculation includes $>150$ molecular, atomic, and ionic species, the species which typically dominate irradiated giant planet atmospheres are H$_2$O, CH$_4$, CO, CO$_2$, NH$_3$, HCN, C$_2$H$_2$, Na, K, TiO, VO, and H$^-$, besides H$_2$ and He \citep{Burrows1999,Madhusudhan2016,Arcangeli2018}. We therefore include opacity due to these species in our atmospheric models, using cross-sections calculated as in \citet{Gandhi2017} and \citet{Gandhi2020} as well as line transition data from a variety of sources (see also \citealt{Piette2020}), which include the HITEMP, HITRAN, and ExoMol databases (H$_2$O, CO and CO$_2$: \citealt{Rothman2010}, CH$_4$: \citealt{Yurchenko2013,Yurchenko2014a}, C$_2$H$_2$: \citealt{Rothman2013,Gordon2017}, NH$_3$: \citealt{Yurchenko2011}, HCN: \citealt{Harris2006,Barber2014}, TiO: \citealt{McKemmish2019}, VO: \citealt{McKemmish2016}, H$_2$-H$_2$ and H$_2$-He Collision-Induced Absorption: \citealt{Richard2012}). The Na and K opacities are calculated as in \citet{Burrows2003} and \citet{Gandhi2017}, while the H$^-$ bound-free and free-free cross sections are calculated using the prescriptions of \citet{Bell1987} and \citet{John1988} (see also \citealt{Arcangeli2018,Parmentier2018,Gandhi2020}).

We consider the stellar irradiation with a flux distribution factor on the day side of $f=2/3$ \citep[e.g.,][]{Burrows2008a}. Given the extreme irradiation received by KELT-1~b, a low day-night energy redistribution efficiency is expected \citep{Cowan2011}. Therefore, we chose to explore models in the limit of no day-night energy redistribution for simplicity. We note, however, that recent studies find that H$_2$ dissociation and recombination may increase heat transport in highly-irradiated atmospheres \citep{Bell2018,Tan2019,Mansfield2020}. 
We also account for the internal heat flux emanating from KELT-1b, which is characterized by the internal temperature, $T_\mathrm{int}$. In the absence of incident irradiation, for instance, on the night side, $T_\mathrm{int}$ is equivalent to the effective temperature used in stellar contexts. While the day-side temperature structure and spectrum are dominated by the irradiation, the night side is dominated by the internal flux and energy advected from the day side. Given the small number of data points available, we do not perform a full atmospheric retrieval which would have more free parameters than data points. Rather, we report physically plausible self-consistent models that provide a consistent explanation for both the day side and night side data. In order to fit the observed data, we explore models with different $T_\mathrm{int}$, metallicities and C/O ratios, as described below. 

We are able to fit the day side data using irradiated models with no redistribution (upper panel of Figure \ref{fig:atm_models}). At the temperatures present in the day-side atmosphere, CO is expected to be a significant opacity source in the IRAC~2 band \citep{Burrows1999,Madhusudhan2016}. However, since the brightness temperature in the IRAC~2 band is comparable to the IRAC~1 band and lower than the TESS band, this suggests the absence of a thermal inversion in the infrared photosphere, which is consistent with the findings of previous studies on KELT-1~b \citep{Beatty2017}. A thermal inversion would instead cause a strong CO emission feature in the IRAC~2 band, which is not seen. In turn, a non-inverted temperature profile corresponds to a low optical/infrared opacity ratio \citep[e.g.,][]{Hubeny2003,Guillot2010}, for example, due to a depletion of optical absorbers such as TiO and VO. Such a depletion may be caused by the gravitational settling of TiO and VO due to their high mean-molecular weight \citep{Spiegel2009} or increased efficiency of cold trap processes \citep{Parmentier2013,Beatty2017}. We find that a model with solar metallicity and C/O=0.5 provides a good fit to the data if the abundances of TiO and VO are just 5\% of those predicted under thermochemical equilibrium (within 1.4$\sigma$, 0.5$\sigma$, 0.8$\sigma$ of the TESS, IRAC~1, and IRAC~2 points, respectively, see red spectrum in the top panel of Figure \ref{fig:atm_models}). The continuum of the model spectrum is close to a blackbody spectrum at $\sim$3000~K, with the decreasing temperature gradient resulting in a slight CO absorption feature in the IRAC~2 band. 

An alternative scenario that involves depleted TiO and VO is that featuring a high C/O ratio \citep{Madhusudhan2011}. We find that a model with solar metallicity in all elements except carbon and C/O=1 is able to fit the TESS, IRAC~1, and IRAC~2 data within 1.5$\sigma$, 1.7$\sigma,$ and 1.9$\sigma$, respectively (blue spectrum in the top panel of figure \ref{fig:atm_models}). The photosphere in the $\sim$0.6-5~$\mu$m range lies between $\sim$0.1-10~bar, where the temperature profile is relatively isothermal. As a result, the emergent thermal emission spectrum is also close to a blackbody at $\sim$3000~K. In this model, the atmospheric chemistry is still dominated by CO, which results in some absorption in the IRAC~2 band. We note that given the deep photosphere in KELT-1~b, thermal dissociation is less prevalent in the photosphere compared to lower-gravity hot Jupiters with shallower photospheres. We also note that, given the strong stellar irradiation, the day-side spectrum of KELT-1~b is not sensitive to internal temperatures below $\gtrsim 2000$~K. As a result, $T_\mathrm{int}$ is constrained only by the night side data and we assume a value of $T_\mathrm{int}$=1500~K for the day side model.

The day side data can be explained by thermal emission alone, without the need for a high albedo, although the latter cannot be ruled out with current data. Figure \ref{fig:atm_models} also shows the T$_{\rm int}$=1000~K cloudy model from \citet{Beatty2020} (from their Figure 6) for reference. Our clear model is able to fit the TESS point slightly better than the \citet{Beatty2020} cloudy model, while the fit to the Spitzer data cannot be compared as their model is truncated at 2.5~$\mu$m. Since the current data do not rule out a cloud-free atmosphere, further observations are needed to constrain whether clouds are present on the day side of KELT-1~b. Indeed, a plausible solution may be a partially cloudy atmosphere, with contributions from both clear and cloudy regions. Future spectroscopic secondary eclipse or phase curve observations may provide more confident constraints on the presence of clouds in KELT-1~b's day side.

We find that the Spitzer night-side brightness temperatures are consistent with a blackbody in the range $\sim$1000-1500~K (within 1$\sigma$). Within the limit of no day-night redistribution, the data can be explained by non-irradiated self-consistent models with $T_\mathrm{int}$ in the range $\sim$1500-2000~K  (lower panel of Figure \ref{fig:atm_models}). This places an upper limit of $\sim$2000~K on $T_\mathrm{int}$, as a smaller internal flux is permitted by the data if additional flux is advected from the day side to the night side. In the models shown in Figure \ref{fig:atm_models}, the absorption features in the 3.6~$\mu$m and 4.5~$\mu$m bands are due to CH$_4$ and CO, respectively. We assume a C/O ratio of 0.5, consistent with the day side model shown in red in the top panel of Figure \ref{fig:atm_models}. However, we note that the night side data are not able to constrain chemical abundances and a range of C/O ratios and metallicities are consistent with the night side observations. Furthermore, we note that since these models do not include energy advected from the day side, solutions with lower $T_\mathrm{int}$ and greater day-night energy redistribution are also possible.

\subsection{Tidal spin-up of the host star}

\cite{Siverd2012} formulated the suggestion of a complete synchronization of the KELT-1 brown dwarf-star system, including the stellar rotation spin-up by tides to a value equal to the orbital period of KELT-1b. This suggestion was backed by the theoretical calculation of a short synchronization timescale \citep{Siverd2012,Matsumura2010} and the large value of measured stellar rotation $v\mathrm{sin}i$ of 56~km/s \citep{Siverd2012}. In this work, we derived a value for the stellar variability period of $P_\mathrm{spots} = 1.52 \pm 0.29$~d (Section~\ref{sec:activity}). When interpreted as the stellar rotation period, it confirms a significant tidal spin-up of the host star, since field stars of mid F spectral type are expected to rotate with a period between 10 and 15~days at the age of KELT-1 of 1.5 to 2~Gyr \citep{Barnes2010,Barnes2016}. Despite evidence of being spun-up, our measurement of stellar variability indicates a slightly longer stellar spin period than the brown dwarf orbital period; thus the process of stellar rotation spin-up might still be ongoing until it reaches full synchronization. However, the uncertainty on the stellar rotational period is only 1-$\sigma$ away from the period of KELT-1b, so they might actually be synchronized at present. We can only speculate if the slight discrepancy between stellar variability period and orbital period might also be caused either by systematic features in the PDC photometry, by difficulties in measuring stellar rotation periods from irregular and low-amplitude photometric variability \citep{Shapiro2020}, or by stellar differential rotation.
We note that the independent study of \cite{Beatty2020}, in analyzing the same TESS data set modeled the photometric variation (which we interpreted as rotational modulation) differently. The authors considered this variation as correlated noise in the TESS data and approximated it with time-dependent polynomials on the order up to seven in different segments of the data. Because of the low amplitude of the signal of 0.02\,\% and potential deviations from strict periodicity due to a time evolution of magnetic surface elements, it is not a trivial task to distinguish whether the source of the flux variation is astrophysical or instrumental in nature. However, the derived planet phase curve parameters are in decent agreement between \cite{Beatty2020} and our work; thus, they appear to be robust against different versions of modeling of the additional flux variation present in the data.

\subsection{Agreement of transit depth between optical and near-IR wavelengths}

The limb darkening corrected transit depth, $(R_p/R_s)^2$ at optical wavelengths of this work agrees with the joint fit of the Spitzer 3.6 and 4.5~$\mu$m data of \cite{Beatty2019} to 0.2$\pm$1.0\,\%. The investigation of a potential wavelength dependence of $(R_p/R_s)^2$ is known as transmission spectroscopy in exoplanet science and is used to characterize the atmospheres of transiting planets. For favorable targets, the transit depth has been found to vary over wavelength by 5\,\% and more \citep{vonEssen2019AlO,Sing2016}. However, KELT-1b is not suited to the atmospheric characterization by transmission spectroscopy because of its large surface gravity and subsequent small value of atmospheric scale height. A typical signal in the transmission spectrum of five scale heights amplitude amounts to a variation in $(R_p/R_s)^2$ of only 0.2\,\%, which therefore would not be detectable for KELT-1b with the current instruments.

However, the sub-percent agreement of $(R_p/R_s)^2$ between the optical and near-IR is helpful for this study because it indicates that our analysis does not miss a significant source of third light in the large TESS aperture, since the modification of the parameters of KELT-1b by third light shows usually a strong wavelength dependence \citep{Mallonn2016,vonEssen2020}. The uncertainty of 1\% does not allow to distinguish between the slightly different values of third light contamination applied in the PDC TESS photometry and  which we derived independently (see Section \ref{sec:third_light}). Their sub-percent difference turns out to be not significant for the currently achieved measurement precision. Also, the agreement of $(R_p/R_s)^2$ over wavelength rules out a significant color-dependent modification of the transit light curve by temperature inhomogeneities on the stellar surface \citep[e.g.,][]{Oshagh2014,Mallonn2018} and confirms the rather low activity level of the star found by the low amplitude of photometric stellar variability in Section~\ref{sec:activity}. In a recent study on hot Jupiter transmission spectroscopy, \cite{Alexoudi2018} described the emergence of an apparent wavelength-variation of $(R_p/R_s)^2$ when the impact factor $b$ applied in the fit deviates from the correct one. The value of $b$ derived here is nearly identical to the value derived in \cite{Beatty2019}. If both values were systematically inaccurate, we would expect a deviation in $(R_p/R_s)^2$. Thus, the precise agreement of our optical value with the near-IR value of \cite{Beatty2019} also indicates a good accuracy of the derived $b$ value of KELT-1b. 

\section{Conclusion}
\label{sec:conclusion}

In this work, we present the detection and analysis of the phase curve of KELT-1b at optical wavelengths, analyzing data taken by the Transiting Exoplanet Survey Satellite (TESS) during cycle 2 and sector 17. Due to the high mass of KELT-1b, there are visible distortions in the TESS light curve caused by the gravitational influence that the brown dwarf companion exerts over the star. First, we modeled the data by combining five components, namely: the primary transit, the secondary eclipse, the phase curve that assumes both reflected light and thermal emission, the Doppler beaming, and the ellipsoidal variation models. From our initial fit, we found evidence of variability in the shape of the phase curve from orbit to orbit. Before jumping to the conclusion that we are observing weather patterns in the atmosphere of KELT-1b, we analyzed TESS photometry in more detail and detect what appears to be stellar activity. After adding this component to our model, we re-fit the data and determined the secondary eclipse depth in the TESS bandpass to be \mbox{304 $\pm$ 75} parts-per-million (ppm). 
Furthermore, we determined the amplitude of the phase curve to be \mbox{129 $\pm$ 27 ppm,} with a corresponding eastward offset between the region of maximum brightness and the substellar point of 19.2 $\pm$ 9.6 degrees; the latter showing good agreement with Spitzer measurements. We determine the day-side brightness temperature in the TESS bandpass of \mbox{3201 $\pm$ 147 K}, slightly higher than what is inferred from Spitzer 3.6 and 4.5 $\mu$m data. 

Using the error-weighted mean of the Spitzer and TESS brightness temperatures, we computed the Bond albedo of KELT-1b to be \mbox{$A_B = 7.1 \pm 7.7$\%}, consistent with zero at 1-$\sigma$, and a heat redistribution efficiency of \mbox{$\epsilon = 0.052 \pm 0.006$}. If the TESS eclipse depth is due to both reflected light and thermal emission, then we derive a geometric albedo of \mbox{$A_g = 25 \pm 11$\%}, greater than the geometric albedos of most other hot Jupiters. For this exercise, we are effectively assuming that the day-side brightness temperature in the TESS band is the error weighted mean of the Spitzer and TESS brightness temperatures, so the high geometric albedo inferred might, in fact, be caused by a non-blackbody spectrum, rather than reflected light. 

Taking the Bond and geometric albedos at face value, we then asked whether they are self-consistent. We estimated the fraction of incident stellar light in the TESS bandpass: 31\%. The minimum Bond albedo should be the product of the geometric albedo and the fraction of incident stellar light in the TESS bandbass, or 7.6\%. This is consistent with the Bond albedo inferred from day and night temperatures, but requires that the geometric albedo is nearly zero outside of the TESS bandpass.

The published H and K-band eclipse measurement suggest a significantly higher day-side temperature than either the TESS of Spitzer measurements. We attribute this to unmodeled  ellipsoidal variation, which would appear as a deeper eclipse depth in observations with insufficient phase coverage. We therefore opt to ignore eclipse depths based on  incomplete phase coverage, and obtain a day-side effective temperature of \mbox{$T_\mathrm{eff} = 2946 \pm 71$~K}, with an improvement of $\chi^2_{\rm red}$ from 6.8 to 1.6. 

A self-consistent 1D radiative transfer model can explain the day-side fluxes from both TESS and Spitzer with thermal emission alone without invoking reflected light. The somewhat lower brightness temperature in the Spitzer IRAC 2 band compared to that in the TESS band can be explained by CO absorption due to a non-inverted temperature profile. 

The Spitzer night-side brightness temperatures place an upper limit of $\sim$2000~K on the internal temperature of KELT-1~b. In the limit of no day-night redistribution, the night side data can be explained using an atmospheric model with an internal temperature of $\sim$1500-2000~K, somewhat higher than the $\sim$850 K internal temperature expected for the brown dwarf \citep[see][]{Beatty2019} if it were unaffected by the stellar irradiation and if the age of the host star is accurate. However, we note that some of the night-side flux which is being attributed to internal flux could potentially be contributed by day-night energy redistribution, which is not included explicitly in the night side model.

The internal temperature of KELT-1~b may be linked to its inflated radius, considering that the radius for such a brown dwarf would be expected to be below 1 R$_J$ if it were isolated and $\approx$2 Gyr old \citep{Burrows1997,Baraffe2003,Siverd2012}. This inflated radius may be a result of an elevated entropy in its interior and a high $T_{\rm int}$. Such a situation is possible if the brown dwarf is young ($\approx10^{8.5}$ yr) and/or if the irradiation has delayed its cooling significantly; in light of this, it would be worth revisiting the age of the host star, carefully accounting for tidal effects and utilizing the rotational activity-age relationship \citep{Mamajek2008}. As such, KELT-1b could provide a valuable testing ground for theories of inflated radii of highly irradiated giant planets.

During the review process of the manuscript of this work, \cite{Beatty2020} published an independent investigation of the same TESS photometry data of KELT-1b. Their derived model parameter values for the secondary eclipse and phase curve variations are broadly consistent with the results presented here.

\section*{Acknowledgements}
We thank the anonymous referee for the helpful comments, and constructive remarks on this manuscript. CvE and GT acknowledge support from the European Social Fund (project No. 09.3.3-LMT-K-712-01-0103) under grant agreement with the Lithuanian Science Council (LMTLT). Funding for the Stellar Astrophysics Centre is provided by The Danish National Research Foundation (Grant agreement No.: DNRF106). This work was supported by a research grant (00028173) from VILLUM FONDEN. GT acknowledges Europlanet 2024 RI project funded by the European Union's Horizon 2020 Research and Innovation Programme (Grant agreement No. 871149). This work made use of PyAstronomy, the SIMBAD data base and VizieR catalog access tool, operated at CDS, Strasbourg, France, and of the NASA Astrophysics Data System (ADS).

\bibliographystyle{aa}
\bibliography{main}

\begin{thebibliography}{116}
\expandafter\ifx\csname natexlab\endcsname\relax\def\natexlab#1{#1}\fi

\bibitem[{{Alexoudi} {et~al.}(2018){Alexoudi}, {Mallonn}, {von Essen},
  {Turner}, {Keles}, {Southworth}, {Mancini}, {Ciceri}, {Granzer}, {Denker},
  {Dineva}, \& {Strassmeier}}]{Alexoudi2018}
{Alexoudi}, X., {Mallonn}, M., {von Essen}, C., {et~al.} 2018, \aap, 620, A142

\bibitem[{{Alonso}(2018)}]{Alonso2018}
{Alonso}, R. 2018, {Characterization of Exoplanets: Secondary Eclipses}, 40

\bibitem[{{Arcangeli} {et~al.}(2018){Arcangeli}, {D{\'e}sert}, {Line}, {Bean},
  {Parmentier}, {Stevenson}, {Kreidberg}, {Fortney}, {Mansfield}, \&
  {Showman}}]{Arcangeli2018}
{Arcangeli}, J., {D{\'e}sert}, J.-M., {Line}, M.~R., {et~al.} 2018, \apjl, 855,
  L30

\bibitem[{{Armstrong} {et~al.}(2016){Armstrong}, {de Mooij}, {Barstow},
  {Osborn}, {Blake}, \& {Saniee}}]{Armstrong2016}
{Armstrong}, D.~J., {de Mooij}, E., {Barstow}, J., {et~al.} 2016, Nature
  Astronomy, 1, 0004

\bibitem[{{Baraffe} {et~al.}(2003){Baraffe}, {Chabrier}, {Barman}, {Allard}, \&
  {Hauschildt}}]{Baraffe2003}
{Baraffe}, I., {Chabrier}, G., {Barman}, T.~S., {Allard}, F., \& {Hauschildt},
  P.~H. 2003, \aap, 402, 701

\bibitem[{{Barber} {et~al.}(2014){Barber}, {Strange}, {Hill}, {Polyansky},
  {Mellau}, {Yurchenko}, \& {Tennyson}}]{Barber2014}
{Barber}, R.~J., {Strange}, J.~K., {Hill}, C., {et~al.} 2014, \mnras, 437, 1828

\bibitem[{{Barnes}(2010)}]{Barnes2010}
{Barnes}, S.~A. 2010, \apj, 722, 222

\bibitem[{{Barnes} {et~al.}(2016){Barnes}, {Weingrill}, {Fritzewski},
  {Strassmeier}, \& {Platais}}]{Barnes2016}
{Barnes}, S.~A., {Weingrill}, J., {Fritzewski}, D., {Strassmeier}, K.~G., \&
  {Platais}, I. 2016, \apj, 823, 16

\bibitem[{{Beatty} {et~al.}(2014){Beatty}, {Collins}, {Fortney}, {Knutson},
  {Gaudi}, {Bruns}, {Showman}, {Eastman}, {Pepper}, {Siverd}, {Stassun}, \&
  {Kielkopf}}]{Beatty2014}
{Beatty}, T.~G., {Collins}, K.~A., {Fortney}, J., {et~al.} 2014, \apj, 783, 112

\bibitem[{{Beatty} {et~al.}(2017){Beatty}, {Madhusudhan}, {Pogge}, {Chung},
  {Bierlya}, {Gaudi}, \& {Latham}}]{Beatty2017}
{Beatty}, T.~G., {Madhusudhan}, N., {Pogge}, R., {et~al.} 2017, \aj, 154, 242

\bibitem[{{Beatty} {et~al.}(2019){Beatty}, {Marley}, {Gaudi}, {Col{\'o}n},
  {Fortney}, \& {Showman}}]{Beatty2019}
{Beatty}, T.~G., {Marley}, M.~S., {Gaudi}, B.~S., {et~al.} 2019, \aj, 158, 166

\bibitem[{{Beatty} {et~al.}(2020){Beatty}, {Wong}, {Fetherolf}, {Line},
  {Shporer}, {Stassun}, {Ricker}, {Seager}, {Winn}, {Jenkins}, {Louie},
  {Schlieder}, {Sha}, {Tenenbaum}, \& {Yahalomi}}]{Beatty2020}
{Beatty}, T.~G., {Wong}, I., {Fetherolf}, T., {et~al.} 2020, \aj, 160, 211

\bibitem[{{Bell} \& {Berrington}(1987)}]{Bell1987}
{Bell}, K.~L. \& {Berrington}, K.~A. 1987, Journal of Physics B Atomic
  Molecular Physics, 20, 801

\bibitem[{{Bell} \& {Cowan}(2018)}]{Bell2018}
{Bell}, T.~J. \& {Cowan}, N.~B. 2018, \apjl, 857, L20

\bibitem[{{Bell} {et~al.}(2019){Bell}, {Zhang}, {Cubillos}, {Dang}, {Fossati},
  {Todorov}, {Cowan}, {Deming}, {Zellem}, {Stevenson}, {Crossfield},
  {Dobbs-Dixon}, {Fortney}, {Knutson}, \& {Line}}]{Bell2019}
{Bell}, T.~J., {Zhang}, M., {Cubillos}, P.~E., {et~al.} 2019, \mnras, 489, 1995

\bibitem[{{Belokurov} {et~al.}(2020){Belokurov}, {Penoyre}, {Oh}, {Iorio},
  {Hodgkin}, {Evans}, {Everall}, {Koposov}, {Tout}, {Izzard}, {Clarke}, \&
  {Brown}}]{Belokurov2020}
{Belokurov}, V., {Penoyre}, Z., {Oh}, S., {et~al.} 2020, \mnras, 496, 1922

\bibitem[{{Bourrier} {et~al.}(2020){Bourrier}, {Kitzmann}, {Kuntzer},
  {Nascimbeni}, {Lendl}, {Lavie}, {Hoeijmakers}, {Pino}, {Ehrenreich}, {Heng},
  {Allart}, {Cegla}, {Dumusque}, {Melo}, {Astudillo-Defru}, {Caldwell},
  {Cretignier}, {Giles}, {Henze}, {Jenkins}, {Lovis}, {Murgas}, {Pepe},
  {Ricker}, {Rose}, {Seager}, {Segransan}, {Su{\'a}rez-Mascare{\~n}o}, {Udry},
  {Vanderspek}, \& {Wyttenbach}}]{Bourrier2019}
{Bourrier}, V., {Kitzmann}, D., {Kuntzer}, T., {et~al.} 2020, \aap, 637, A36

\bibitem[{{Burrows} {et~al.}(2008{\natexlab{a}}){Burrows}, {Budaj}, \&
  {Hubeny}}]{Burrows2008b}
{Burrows}, A., {Budaj}, J., \& {Hubeny}, I. 2008{\natexlab{a}}, \apj, 678, 1436

\bibitem[{{Burrows} {et~al.}(2008{\natexlab{b}}){Burrows}, {Budaj}, \&
  {Hubeny}}]{Burrows2008a}
{Burrows}, A., {Budaj}, J., \& {Hubeny}, I. 2008{\natexlab{b}}, \apj, 678, 1436

\bibitem[{{Burrows} {et~al.}(1997){Burrows}, {Marley}, {Hubbard}, {Lunine},
  {Guillot}, {Saumon}, {Freedman}, {Sudarsky}, \& {Sharp}}]{Burrows1997}
{Burrows}, A., {Marley}, M., {Hubbard}, W.~B., {et~al.} 1997, \apj, 491, 856

\bibitem[{{Burrows} \& {Sharp}(1999)}]{Burrows1999}
{Burrows}, A. \& {Sharp}, C.~M. 1999, \apj, 512, 843

\bibitem[{{Burrows} \& {Volobuyev}(2003)}]{Burrows2003}
{Burrows}, A. \& {Volobuyev}, M. 2003, \apj, 583, 985

\bibitem[{{Carter} \& {Winn}(2009)}]{Carter2009}
{Carter}, J.~A. \& {Winn}, J.~N. 2009, \apj, 704, 51

\bibitem[{Charbonneau \& Deming(2007)}]{Charbonneau2007}
Charbonneau, D. \& Deming, D. 2007, arXiv preprint arXiv:0706.1047

\bibitem[{{Claret}(2017)}]{Claret2017}
{Claret}, A. 2017, \aap, 600, A30

\bibitem[{{Claret} \& {Bloemen}(2011)}]{claret2011}
{Claret}, A. \& {Bloemen}, S. 2011, \aap, 529, A75

\bibitem[{{Cowan} \& {Agol}(2008)}]{Cowan2008}
{Cowan}, N.~B. \& {Agol}, E. 2008, \apjl, 678, L129

\bibitem[{{Cowan} \& {Agol}(2011)}]{Cowan2011}
{Cowan}, N.~B. \& {Agol}, E. 2011, \apj, 729, 54

\bibitem[{{Croll} {et~al.}(2015){Croll}, {Albert}, {Jayawardhana}, {Cushing},
  {Moutou}, {Lafreniere}, {Johnson}, {Bonomo}, {Deleuil}, \&
  {Fortney}}]{Croll2015}
{Croll}, B., {Albert}, L., {Jayawardhana}, R., {et~al.} 2015, \apj, 802, 28

\bibitem[{{Daylan} {et~al.}(2019){Daylan}, {G{\"u}nther}, {Mikal-Evans},
  {Sing}, {Wong}, {Shporer}, {Crossfield}, {Niraula}, {de Wit}, {Koll},
  {Parmentier}, {Fetherolf}, {Kane}, {Ricker}, {Vand erspek}, {Seager}, {Winn},
  {Jenkins}, {Caldwell}, {Charbonneau}, {Henze}, {Paegert}, {Rinehart}, {Rose},
  {Sha}, {Quintana}, \& {Villasenor}}]{Daylan2019}
{Daylan}, T., {G{\"u}nther}, M.~N., {Mikal-Evans}, T., {et~al.} 2019, arXiv
  e-prints, arXiv:1909.03000

\bibitem[{{Deming} {et~al.}(2019){Deming}, {Louie}, \& {Sheets}}]{Deming2019}
{Deming}, D., {Louie}, D., \& {Sheets}, H. 2019, \pasp, 131, 013001

\bibitem[{{Demory} {et~al.}(2011){Demory}, {Seager}, {Madhusudhan}, {Kjeldsen},
  {Christensen-Dalsgaard}, {Gillon}, {Rowe}, {Welsh}, {Adams}, {Dupree},
  {McCarthy}, {Kulesa}, {Borucki}, \& {Koch}}]{Demory2011}
{Demory}, B.-O., {Seager}, S., {Madhusudhan}, N., {et~al.} 2011, \apjl, 735,
  L12

\bibitem[{{Esteves} {et~al.}(2015){Esteves}, {De Mooij}, \&
  {Jayawardhana}}]{Esteves2015}
{Esteves}, L.~J., {De Mooij}, E. J.~W., \& {Jayawardhana}, R. 2015, \apj, 804,
  150

\bibitem[{{Evans} {et~al.}(2016){Evans}, {Sing}, {Wakeford}, {Nikolov},
  {Ballester}, {Drummond}, {Kataria}, {Gibson}, {Amundsen}, \&
  {Spake}}]{Evans2016}
{Evans}, T.~M., {Sing}, D.~K., {Wakeford}, H.~R., {et~al.} 2016, \apjl, 822, L4

\bibitem[{{Gaia Collaboration} {et~al.}(2018){Gaia Collaboration}, {Brown},
  {Vallenari}, {Prusti}, {de Bruijne}, {Babusiaux}, {Bailer-Jones}, {Biermann},
  {Evans}, {Eyer}, {Jansen}, {Jordi}, {Klioner}, {Lammers}, {Lindegren},
  {Luri}, {Mignard}, {Panem}, {Pourbaix}, {Randich}, {Sartoretti}, {Siddiqui},
  {Soubiran}, {van Leeuwen}, {Walton}, {Arenou}, {Bastian}, {Cropper},
  {Drimmel}, {Katz}, {Lattanzi}, {Bakker}, {Cacciari}, {Casta{\~n}eda},
  {Chaoul}, {Cheek}, {De Angeli}, {Fabricius}, {Guerra}, {Holl}, {Masana},
  {Messineo}, {Mowlavi}, {Nienartowicz}, {Panuzzo}, {Portell}, {Riello},
  {Seabroke}, {Tanga}, {Th{\'e}venin}, {Gracia-Abril}, {Comoretto},
  {Garcia-Reinaldos}, {Teyssier}, {Altmann}, {Andrae}, {Audard},
  {Bellas-Velidis}, {Benson}, {Berthier}, {Blomme}, {Burgess}, {Busso},
  {Carry}, {Cellino}, {Clementini}, {Clotet}, {Creevey}, {Davidson}, {De
  Ridder}, {Delchambre}, {Dell'Oro}, {Ducourant},
  {Fern{\'a}ndez-Hern{\'a}ndez}, {Fouesneau}, {Fr{\'e}mat}, {Galluccio},
  {Garc{\'\i}a-Torres}, {Gonz{\'a}lez-N{\'u}{\~n}ez}, {Gonz{\'a}lez-Vidal},
  {Gosset}, {Guy}, {Halbwachs}, {Hambly}, {Harrison}, {Hern{\'a}ndez},
  {Hestroffer}, {Hodgkin}, {Hutton}, {Jasniewicz}, {Jean-Antoine-Piccolo},
  {Jordan}, {Korn}, {Krone-Martins}, {Lanzafame}, {Lebzelter}, {L{\"o}ffler},
  {Manteiga}, {Marrese}, {Mart{\'\i}n-Fleitas}, {Moitinho}, {Mora}, {Muinonen},
  {Osinde}, {Pancino}, {Pauwels}, {Petit}, {Recio-Blanco}, {Richards},
  {Rimoldini}, {Robin}, {Sarro}, {Siopis}, {Smith}, {Sozzetti}, {S{\"u}veges},
  {Torra}, {van Reeven}, {Abbas}, {Abreu Aramburu}, {Accart}, {Aerts},
  {Altavilla}, {{\'A}lvarez}, {Alvarez}, {Alves}, {Anderson}, {Andrei},
  {Anglada Varela}, {Antiche}, {Antoja}, {Arcay}, {Astraatmadja}, {Bach},
  {Baker}, {Balaguer-N{\'u}{\~n}ez}, {Balm}, {Barache}, {Barata}, {Barbato},
  {Barblan}, {Barklem}, {Barrado}, {Barros}, {Barstow}, {Bartholom{\'e}
  Mu{\~n}oz}, {Bassilana}, {Becciani}, {Bellazzini}, {Berihuete}, {Bertone},
  {Bianchi}, {Bienaym{\'e}}, {Blanco-Cuaresma}, {Boch}, {Boeche}, {Bombrun},
  {Borrachero}, {Bossini}, {Bouquillon}, {Bourda}, {Bragaglia}, {Bramante},
  {Breddels}, {Bressan}, {Brouillet}, {Br{\"u}semeister}, {Brugaletta},
  {Bucciarelli}, {Burlacu}, {Busonero}, {Butkevich}, {Buzzi}, {Caffau},
  {Cancelliere}, {Cannizzaro}, {Cantat-Gaudin}, {Carballo}, {Carlucci},
  {Carrasco}, {Casamiquela}, {Castellani}, {Castro-Ginard}, {Charlot},
  {Chemin}, {Chiavassa}, {Cocozza}, {Costigan}, {Cowell}, {Crifo}, {Crosta},
  {Crowley}, {Cuypers}, {Dafonte}, {Damerdji}, {Dapergolas}, {David}, {David},
  {de Laverny}, {De Luise}, {De March}, {de Martino}, {de Souza}, {de Torres},
  {Debosscher}, {del Pozo}, {Delbo}, {Delgado}, {Delgado}, {Di Matteo},
  {Diakite}, {Diener}, {Distefano}, {Dolding}, {Drazinos}, {Dur{\'a}n},
  {Edvardsson}, {Enke}, {Eriksson}, {Esquej}, {Eynard Bontemps}, {Fabre},
  {Fabrizio}, {Faigler}, {Falc{\~a}o}, {Farr{\`a}s Casas}, {Federici},
  {Fedorets}, {Fernique}, {Figueras}, {Filippi}, {Findeisen}, {Fonti},
  {Fraile}, {Fraser}, {Fr{\'e}zouls}, {Gai}, {Galleti}, {Garabato},
  {Garc{\'\i}a-Sedano}, {Garofalo}, {Garralda}, {Gavel}, {Gavras}, {Gerssen},
  {Geyer}, {Giacobbe}, {Gilmore}, {Girona}, {Giuffrida}, {Glass}, {Gomes},
  {Granvik}, {Gueguen}, {Guerrier}, {Guiraud}, {Guti{\'e}rrez-S{\'a}nchez},
  {Haigron}, {Hatzidimitriou}, {Hauser}, {Haywood}, {Heiter}, {Helmi}, {Heu},
  {Hilger}, {Hobbs}, {Hofmann}, {Holland}, {Huckle}, {Hypki}, {Icardi},
  {Jan{\ss}en}, {Jevardat de Fombelle}, {Jonker}, {Juh{\'a}sz}, {Julbe},
  {Karampelas}, {Kewley}, {Klar}, {Kochoska}, {Kohley}, {Kolenberg},
  {Kontizas}, {Kontizas}, {Koposov}, {Kordopatis}, {Kostrzewa-Rutkowska},
  {Koubsky}, {Lambert}, {Lanza}, {Lasne}, {Lavigne}, {Le Fustec}, {Le
  Poncin-Lafitte}, {Lebreton}, {Leccia}, {Leclerc}, {Lecoeur-Taibi},
  {Lenhardt}, {Leroux}, {Liao}, {Licata}, {Lindstr{\o}m}, {Lister}, {Livanou},
  {Lobel}, {L{\'o}pez}, {Managau}, {Mann}, {Mantelet}, {Marchal}, {Marchant},
  {Marconi}, {Marinoni}, {Marschalk{\'o}}, {Marshall}, {Martino}, {Marton},
  {Mary}, {Massari}, {Matijevi{\v{c}}}, {Mazeh}, {McMillan}, {Messina},
  {Michalik}, {Millar}, {Molina}, {Molinaro}, {Moln{\'a}r}, {Montegriffo},
  {Mor}, {Morbidelli}, {Morel}, {Morris}, {Mulone}, {Muraveva}, {Musella},
  {Nelemans}, {Nicastro}, {Noval}, {O'Mullane}, {Ord{\'e}novic},
  {Ord{\'o}{\~n}ez-Blanco}, {Osborne}, {Pagani}, {Pagano}, {Pailler},
  {Palacin}, {Palaversa}, {Panahi}, {Pawlak}, {Piersimoni}, {Pineau}, {Plachy},
  {Plum}, {Poggio}, {Poujoulet}, {Pr{\v{s}}a}, {Pulone}, {Racero}, {Ragaini},
  {Rambaux}, {Ramos-Lerate}, {Regibo}, {Reyl{\'e}}, {Riclet}, {Ripepi}, {Riva},
  {Rivard}, {Rixon}, {Roegiers}, {Roelens}, {Romero-G{\'o}mez}, {Rowell},
  {Royer}, {Ruiz-Dern}, {Sadowski}, {Sagrist{\`a} Sell{\'e}s}, {Sahlmann},
  {Salgado}, {Salguero}, {Sanna}, {Santana-Ros}, {Sarasso}, {Savietto},
  {Schultheis}, {Sciacca}, {Segol}, {Segovia}, {S{\'e}gransan}, {Shih},
  {Siltala}, {Silva}, {Smart}, {Smith}, {Solano}, {Solitro}, {Sordo}, {Soria
  Nieto}, {Souchay}, {Spagna}, {Spoto}, {Stampa}, {Steele},
  {Steidelm{\"u}ller}, {Stephenson}, {Stoev}, {Suess}, {Surdej}, {Szabados},
  {Szegedi-Elek}, {Tapiador}, {Taris}, {Tauran}, {Taylor}, {Teixeira},
  {Terrett}, {Teyssand ier}, {Thuillot}, {Titarenko}, {Torra Clotet}, {Turon},
  {Ulla}, {Utrilla}, {Uzzi}, {Vaillant}, {Valentini}, {Valette}, {van Elteren},
  {Van Hemelryck}, {van Leeuwen}, {Vaschetto}, {Vecchiato}, {Veljanoski},
  {Viala}, {Vicente}, {Vogt}, {von Essen}, {Voss}, {Votruba}, {Voutsinas},
  {Walmsley}, {Weiler}, {Wertz}, {Wevers}, {Wyrzykowski}, {Yoldas},
  {{\v{Z}}erjal}, {Ziaeepour}, {Zorec}, {Zschocke}, {Zucker}, {Zurbach}, \&
  {Zwitter}}]{GaiaCollaboration2018}
{Gaia Collaboration}, {Brown}, A.~G.~A., {Vallenari}, A., {et~al.} 2018, \aap,
  616, A1

\bibitem[{{Gandhi} \& {Madhusudhan}(2017)}]{Gandhi2017}
{Gandhi}, S. \& {Madhusudhan}, N. 2017, \mnras, 472, 2334

\bibitem[{{Gandhi} {et~al.}(2020){Gandhi}, {Madhusudhan}, \&
  {Mandell}}]{Gandhi2020}
{Gandhi}, S., {Madhusudhan}, N., \& {Mandell}, A. 2020, \aj, 159, 232

\bibitem[{{Gordon} {et~al.}(2017){Gordon}, {Rothman}, {Hill}, {Kochanov},
  {Tan}, {Bernath}, {Birk}, {Boudon}, {Campargue}, {Chance}, {Drouin}, {Flaud},
  {Gamache}, {Hodges}, {Jacquemart}, {Perevalov}, {Perrin}, {Shine}, {Smith},
  {Tennyson}, {Toon}, {Tran}, {Tyuterev}, {Barbe}, {Cs{\'a}sz{\'a}r}, {Devi},
  {Furtenbacher}, {Harrison}, {Hartmann}, {Jolly}, {Johnson}, {Karman},
  {Kleiner}, {Kyuberis}, {Loos}, {Lyulin}, {Massie}, {Mikhailenko},
  {Moazzen-Ahmadi}, {M{\"u}ller}, {Naumenko}, {Nikitin}, {Polyansky}, {Rey},
  {Rotger}, {Sharpe}, {Sung}, {Starikova}, {Tashkun}, {Auwera}, {Wagner},
  {Wilzewski}, {Wcis{\l}o}, {Yu}, \& {Zak}}]{Gordon2017}
{Gordon}, I.~E., {Rothman}, L.~S., {Hill}, C., {et~al.} 2017, \jqsrt, 203, 3

\bibitem[{{Guillot}(2010)}]{Guillot2010}
{Guillot}, T. 2010, \aap, 520, A27

\bibitem[{{Guillot} {et~al.}(1996){Guillot}, {Burrows}, {Hubbard}, {Lunine}, \&
  {Saumon}}]{Guillot1996}
{Guillot}, T., {Burrows}, A., {Hubbard}, W.~B., {Lunine}, J.~I., \& {Saumon},
  D. 1996, \apjl, 459, L35

\bibitem[{Göttingen(2018)}]{phoenix}
Göttingen, G.-A.-U. 2018, Göttingen Spectral Library by PHOENIX

\bibitem[{{Harris} {et~al.}(2006){Harris}, {Tennyson}, {Kaminsky}, {Pavlenko},
  \& {Jones}}]{Harris2006}
{Harris}, G.~J., {Tennyson}, J., {Kaminsky}, B.~M., {Pavlenko}, Y.~V., \&
  {Jones}, H.~R.~A. 2006, \mnras, 367, 400

\bibitem[{{Huang} {et~al.}(2018){Huang}, {Burt}, {Vanderburg}, {G{\"u}nther},
  {Shporer}, {Dittmann}, {Winn}, {Wittenmyer}, {Sha}, {Kane}, {Ricker}, {Vand
  erspek}, {Latham}, {Seager}, {Jenkins}, {Caldwell}, {Collins}, {Guerrero},
  {Smith}, {Quinn}, {Udry}, {Pepe}, {Bouchy}, {S{\'e}gransan}, {Lovis},
  {Ehrenreich}, {Marmier}, {Mayor}, {Wohler}, {Haworth}, {Morgan}, {Fausnaugh},
  {Ciardi}, {Christiansen}, {Charbonneau}, {Dragomir}, {Deming}, {Glidden},
  {Levine}, {McCullough}, {Yu}, {Narita}, {Nguyen}, {Morton}, {Pepper},
  {P{\'a}l}, {Rodriguez}, {Stassun}, {Torres}, {Sozzetti}, {Doty},
  {Christensen-Dalsgaard}, {Laughlin}, {Clampin}, {Bean}, {Buchhave}, {Bakos},
  {Sato}, {Ida}, {Kaltenegger}, {Palle}, {Sasselov}, {Butler}, {Lissauer},
  {Ge}, \& {Rinehart}}]{Huang2018}
{Huang}, C.~X., {Burt}, J., {Vanderburg}, A., {et~al.} 2018, \apjl, 868, L39

\bibitem[{{Hubeny} {et~al.}(2003){Hubeny}, {Burrows}, \&
  {Sudarsky}}]{Hubeny2003}
{Hubeny}, I., {Burrows}, A., \& {Sudarsky}, D. 2003, \apj, 594, 1011

\bibitem[{{Jansen} \& {Kipping}(2020)}]{Jansen2020}
{Jansen}, T. \& {Kipping}, D. 2020, \mnras, 494, 4077

\bibitem[{{Jenkins}(2017)}]{Jenkins2017}
{Jenkins}, J.~M. 2017, {Kepler Data Processing Handbook: KSCI-19081-002}, Tech.
  rep.

\bibitem[{{Jenkins} {et~al.}(2016){Jenkins}, {Twicken}, {McCauliff},
  {Campbell}, {Sanderfer}, {Lung}, {Mansouri-Samani}, {Girouard}, {Tenenbaum},
  {Klaus}, {Smith}, {Caldwell}, {Chacon}, {Henze}, {Heiges}, {Latham},
  {Morgan}, {Swade}, {Rinehart}, \& {Vanderspek}}]{Jenkins2016}
{Jenkins}, J.~M., {Twicken}, J.~D., {McCauliff}, S., {et~al.} 2016, Society of
  Photo-Optical Instrumentation Engineers (SPIE) Conference Series, Vol. 9913,
  {The TESS science processing operations center}, 99133E

\bibitem[{{John}(1988)}]{John1988}
{John}, T.~L. 1988, \aap, 193, 189

\bibitem[{{Jones} {et~al.}(2001){Jones}, {Oliphant}, {Peterson},
  {et~al.}}]{Jones2001}
{Jones}, E., {Oliphant}, T., {Peterson}, P., {et~al.} 2001, {SciPy}: Open
  source scientific tools for {Python}, \url{http://www.scipy.org}

\bibitem[{{Keating} \& {Cowan}(2017)}]{Keating2017}
{Keating}, D. \& {Cowan}, N.~B. 2017, \apjl, 849, L5

\bibitem[{{Kreidberg}(2018)}]{Kreidberg2018b}
{Kreidberg}, L. 2018, {Exoplanet Atmosphere Measurements from Transmission
  Spectroscopy and Other Planet Star Combined Light Observations}, 100

\bibitem[{{Kreidberg} {et~al.}(2018){Kreidberg}, {Line}, {Parmentier},
  {Stevenson}, {Louden}, {Bonnefoy}, {Faherty}, {Henry}, {Williamson},
  {Stassun}, {Beatty}, {Bean}, {Fortney}, {Showman}, {D{\'e}sert}, \&
  {Arcangeli}}]{Kreidberg2018}
{Kreidberg}, L., {Line}, M.~R., {Parmentier}, V., {et~al.} 2018, \aj, 156, 17

\bibitem[{{Loeb} \& {Gaudi}(2003)}]{Loeb2003}
{Loeb}, A. \& {Gaudi}, B.~S. 2003, \apjl, 588, L117

\bibitem[{{Lomb}(1976)}]{Lomb}
{Lomb}, N.~R. 1976, \apss, 39, 447

\bibitem[{{Maciejewski} {et~al.}(2018){Maciejewski}, {Fern{\'a}ndez},
  {Aceituno}, {Mart{\'\i}n-Ruiz}, {Ohlert}, {Dimitrov}, {Szyszka}, {von Essen},
  {Mugrauer}, {Bischoff}, {Michel}, {Mallonn}, {Stangret}, \&
  {Mo{\'z}dzierski}}]{Maciejewski2018}
{Maciejewski}, G., {Fern{\'a}ndez}, M., {Aceituno}, F., {et~al.} 2018, \actaa,
  68, 371

\bibitem[{{Madhusudhan}(2019)}]{Madhusudhan2019}
{Madhusudhan}, N. 2019, \araa, 57, 617

\bibitem[{{Madhusudhan} {et~al.}(2016){Madhusudhan}, {Ag{\'u}ndez}, {Moses}, \&
  {Hu}}]{Madhusudhan2016}
{Madhusudhan}, N., {Ag{\'u}ndez}, M., {Moses}, J.~I., \& {Hu}, Y. 2016, \ssr,
  205, 285

\bibitem[{{Madhusudhan} {et~al.}(2011){Madhusudhan}, {Mousis}, {Johnson}, \&
  {Lunine}}]{Madhusudhan2011}
{Madhusudhan}, N., {Mousis}, O., {Johnson}, T.~V., \& {Lunine}, J.~I. 2011,
  \apj, 743, 191

\bibitem[{{Mallonn} {et~al.}(2018){Mallonn}, {Herrero}, {Juvan}, {von Essen},
  {Rosich}, {Ribas}, {Granzer}, {Alexoudi}, \& {Strassmeier}}]{Mallonn2018}
{Mallonn}, M., {Herrero}, E., {Juvan}, I.~G., {et~al.} 2018, \aap, 614, A35

\bibitem[{{Mallonn} {et~al.}(2019){Mallonn}, {K{\"o}hler}, {Alexoudi}, {von
  Essen}, {Granzer}, {Poppenhaeger}, \& {Strassmeier}}]{Mallonn2019}
{Mallonn}, M., {K{\"o}hler}, J., {Alexoudi}, X., {et~al.} 2019, \aap, 624, A62

\bibitem[{{Mallonn} \& {Strassmeier}(2016)}]{Mallonn2016}
{Mallonn}, M. \& {Strassmeier}, K.~G. 2016, \aap, 590, A100

\bibitem[{Mamajek \& Hillenbrand(2008)}]{Mamajek2008}
Mamajek, E.~E. \& Hillenbrand, L.~A. 2008, The Astrophysical Journal, 687, 1264

\bibitem[{{Mandel} \& {Agol}(2002)}]{MandelAgol2002}
{Mandel}, K. \& {Agol}, E. 2002, \apjl, 580, L171

\bibitem[{{Mansfield} {et~al.}(2020){Mansfield}, {Bean}, {Stevenson},
  {Komacek}, {Bell}, {Tan}, {Malik}, {Beatty}, {Wong}, {Cowan}, {Dang},
  {D{\'e}sert}, {Fortney}, {Gaudi}, {Keating}, {Kempton}, {Kreidberg}, {Line},
  {Parmentier}, {Stassun}, {Swain}, \& {Zellem}}]{Mansfield2020}
{Mansfield}, M., {Bean}, J.~L., {Stevenson}, K.~B., {et~al.} 2020, \apjl, 888,
  L15

\bibitem[{{Matsumura} {et~al.}(2010){Matsumura}, {Peale}, \&
  {Rasio}}]{Matsumura2010}
{Matsumura}, S., {Peale}, S.~J., \& {Rasio}, F.~A. 2010, \apj, 725, 1995

\bibitem[{{McKemmish} {et~al.}(2019){McKemmish}, {Masseron}, {Hoeijmakers},
  {Perez-Mesa}, {Grimm}, {Yurchenko}, \& {Tennyson}}]{McKemmish2019}
{McKemmish}, L.~K., {Masseron}, T., {Hoeijmakers}, H.~J., {et~al.} 2019,
  \mnras, 488, 2836

\bibitem[{{McKemmish} {et~al.}(2016){McKemmish}, {Yurchenko}, \&
  {Tennyson}}]{McKemmish2016}
{McKemmish}, L.~K., {Yurchenko}, S.~N., \& {Tennyson}, J. 2016, \mnras, 463,
  771

\bibitem[{{Mislis} \& {Hodgkin}(2012)}]{Mislis2012}
{Mislis}, D. \& {Hodgkin}, S. 2012, \mnras, 422, 1512

\bibitem[{{Morris}(1985)}]{Morris1985}
{Morris}, S.~L. 1985, \apj, 295, 143

\bibitem[{{Mugrauer}(2019)}]{Mugrauer2019}
{Mugrauer}, M. 2019, \mnras, 490, 5088

\bibitem[{{Nielsen} {et~al.}(2020){Nielsen}, {Brahm}, {Bouchy}, {Espinoza},
  {Turner}, {Rappaport}, {Pearce}, {Ricker}, {Vanderspek}, {Latham}, {Seager},
  {Winn}, {Jenkins}, {Acton}, {Bakos}, {Barclay}, {Barkaoui}, {Bhatti},
  {Brice{\~n}o}, {Bryant}, {Burleigh}, {Ciardi}, {Collins}, {Collins}, {Cooke},
  {Csubry}, {dos Santos}, {Eigm{\"u}ller}, {Fausnaugh}, {Gan}, {Gillon},
  {Goad}, {Guerrero}, {Hagelberg}, {Hart}, {Henning}, {Huang}, {Jehin},
  {Jenkins}, {Jord{\'a}n}, {Kielkopf}, {Kossakowski}, {Lavie}, {Law}, {Lendl},
  {de Leon}, {Lovis}, {Mann}, {Marmier}, {McCormac}, {Mori}, {Moyano},
  {Narita}, {Osip}, {Otegi}, {Pepe}, {Pozuelos}, {Raynard}, {Relles}, {Sarkis},
  {S{\'e}gransan}, {Seidel}, {Shporer}, {Stalport}, {Stockdale}, {Suc},
  {Tamura}, {Tan}, {Tilbrook}, {Ting}, {Trifonov}, {Udry}, {Vanderburg},
  {Wheatley}, {Wingham}, {Zhan}, \& {Ziegler}}]{Nielsen2020}
{Nielsen}, L.~D., {Brahm}, R., {Bouchy}, F., {et~al.} 2020, \aap, 639, A76

\bibitem[{{Oshagh} {et~al.}(2014){Oshagh}, {Santos}, {Ehrenreich},
  {Haghighipour}, {Figueira}, {Santerne}, \& {Montalto}}]{Oshagh2014}
{Oshagh}, M., {Santos}, N.~C., {Ehrenreich}, D., {et~al.} 2014, \aap, 568, A99

\bibitem[{{Parmentier} \& {Crossfield}(2018)}]{Parmentier2018}
{Parmentier}, V. \& {Crossfield}, I. J.~M. 2018, {Exoplanet Phase Curves:
  Observations and Theory}, 116

\bibitem[{{Parmentier} {et~al.}(2013){Parmentier}, {Showman}, \&
  {Lian}}]{Parmentier2013}
{Parmentier}, V., {Showman}, A.~P., \& {Lian}, Y. 2013, \aap, 558, A91

\bibitem[{{Patil} {et~al.}(2010){Patil}, {Huard}, \& {Fonnesbeck}}]{Patil2010}
{Patil}, A., {Huard}, D., \& {Fonnesbeck}, C.~J. 2010, Journal of Statistical
  Software, 35, 1

\bibitem[{{Pepper} {et~al.}(2012){Pepper}, {Kuhn}, {Siverd}, {James}, \&
  {Stassun}}]{Pepper2012}
{Pepper}, J., {Kuhn}, R.~B., {Siverd}, R., {James}, D., \& {Stassun}, K. 2012,
  \pasp, 124, 230

\bibitem[{{Piette} {et~al.}(2020){Piette}, {Madhusudhan}, {McKemmish},
  {Gandhi}, {Masseron}, \& {Welbanks}}]{Piette2020}
{Piette}, A. A.~A., {Madhusudhan}, N., {McKemmish}, L.~K., {et~al.} 2020,
  \mnras, 496, 3870

\bibitem[{{Piskorz} {et~al.}(2015){Piskorz}, {Knutson}, {Ngo}, {Muirhead},
  {Batygin}, {Crepp}, {Hinkley}, \& {Morton}}]{Piskorz2015}
{Piskorz}, D., {Knutson}, H.~A., {Ngo}, H., {et~al.} 2015, \apj, 814, 148

\bibitem[{{Richard} {et~al.}(2012){Richard}, {Gordon}, {Rothman}, {Abel},
  {Frommhold}, {Gustafsson}, {Hartmann}, {Hermans}, {Lafferty}, {Orton},
  {Smith}, \& {Tran}}]{Richard2012}
{Richard}, C., {Gordon}, I.~E., {Rothman}, L.~S., {et~al.} 2012, \jqsrt, 113,
  1276

\bibitem[{{Ricker} {et~al.}(2015){Ricker}, {Winn}, {Vanderspek}, {Latham},
  {Bakos}, {Bean}, {Berta-Thompson}, {Brown}, {Buchhave}, {Butler}, {Butler},
  {Chaplin}, {Charbonneau}, {Christensen-Dalsgaard}, {Clampin}, {Deming},
  {Doty}, {De Lee}, {Dressing}, {Dunham}, {Endl}, {Fressin}, {Ge}, {Henning},
  {Holman}, {Howard}, {Ida}, {Jenkins}, {Jernigan}, {Johnson}, {Kaltenegger},
  {Kawai}, {Kjeldsen}, {Laughlin}, {Levine}, {Lin}, {Lissauer}, {MacQueen},
  {Marcy}, {McCullough}, {Morton}, {Narita}, {Paegert}, {Palle}, {Pepe},
  {Pepper}, {Quirrenbach}, {Rinehart}, {Sasselov}, {Sato}, {Seager},
  {Sozzetti}, {Stassun}, {Sullivan}, {Szentgyorgyi}, {Torres}, {Udry}, \&
  {Villasenor}}]{Ricker2015}
{Ricker}, G.~R., {Winn}, J.~N., {Vanderspek}, R., {et~al.} 2015, Journal of
  Astronomical Telescopes, Instruments, and Systems, 1, 014003

\bibitem[{{Rothman} {et~al.}(2013){Rothman}, {Gordon}, {Babikov}, {Barbe},
  {Chris Benner}, {Bernath}, {Birk}, {Bizzocchi}, {Boudon}, {Brown},
  {Campargue}, {Chance}, {Cohen}, {Coudert}, {Devi}, {Drouin}, {Fayt}, {Flaud},
  {Gamache}, {Harrison}, {Hartmann}, {Hill}, {Hodges}, {Jacquemart}, {Jolly},
  {Lamouroux}, {Le Roy}, {Li}, {Long}, {Lyulin}, {Mackie}, {Massie},
  {Mikhailenko}, {M{\"u}ller}, {Naumenko}, {Nikitin}, {Orphal}, {Perevalov},
  {Perrin}, {Polovtseva}, {Richard}, {Smith}, {Starikova}, {Sung}, {Tashkun},
  {Tennyson}, {Toon}, {Tyuterev}, \& {Wagner}}]{Rothman2013}
{Rothman}, L.~S., {Gordon}, I.~E., {Babikov}, Y., {et~al.} 2013, \jqsrt, 130, 4

\bibitem[{{Rothman} {et~al.}(2010){Rothman}, {Gordon}, {Barber}, {Dothe},
  {Gamache}, {Goldman}, {Perevalov}, {Tashkun}, \& {Tennyson}}]{Rothman2010}
{Rothman}, L.~S., {Gordon}, I.~E., {Barber}, R.~J., {et~al.} 2010, \jqsrt, 111,
  2139

\bibitem[{{Rowe} {et~al.}(2008){Rowe}, {Matthews}, {Seager}, {Miller-Ricci},
  {Sasselov}, {Kuschnig}, {Guenther}, {Moffat}, {Rucinski}, {Walker}, \&
  {Weiss}}]{Rowe2008}
{Rowe}, J.~F., {Matthews}, J.~M., {Seager}, S., {et~al.} 2008, \apj, 689, 1345

\bibitem[{{Sackett}(1999)}]{Sackett1999}
{Sackett}, P.~D. 1999, in NATO ASIC Proc. 532: Planets Outside the Solar
  System: Theory and Observations, ed. J.-M. {Mariotti} \& D.~{Alloin}, 189

\bibitem[{{Scargle}(1982)}]{Scargle}
{Scargle}, J.~D. 1982, \apj, 263, 835

\bibitem[{{Schlegel} {et~al.}(1998){Schlegel}, {Finkbeiner}, \&
  {Davis}}]{Schlegel:1998}
{Schlegel}, D.~J., {Finkbeiner}, D.~P., \& {Davis}, M. 1998, \apj, 500, 525

\bibitem[{{Shapiro} {et~al.}(2020){Shapiro}, {Amazo-G{\'o}mez}, {Krivova}, \&
  {Solanki}}]{Shapiro2020}
{Shapiro}, A.~I., {Amazo-G{\'o}mez}, E.~M., {Krivova}, N.~A., \& {Solanki},
  S.~K. 2020, \aap, 633, A32

\bibitem[{{Shporer}(2017)}]{Shporer2017}
{Shporer}, A. 2017, \pasp, 129, 072001

\bibitem[{{Shporer} {et~al.}(2019){Shporer}, {Wong}, {Huang}, {Line},
  {Stassun}, {Fetherolf}, {Kane}, {Bouma}, {Daylan}, {G{\"u}enther}, {Ricker},
  {Latham}, {Vanderspek}, {Seager}, {Winn}, {Jenkins}, {Glidden},
  {Berta-Thompson}, {Ting}, {Li}, \& {Haworth}}]{Shporer2019}
{Shporer}, A., {Wong}, I., {Huang}, C.~X., {et~al.} 2019, \aj, 157, 178

\bibitem[{{Sing} {et~al.}(2016){Sing}, {Fortney}, {Nikolov}, {Wakeford},
  {Kataria}, {Evans}, {Aigrain}, {Ballester}, {Burrows}, {Deming},
  {D{\'e}sert}, {Gibson}, {Henry}, {Huitson}, {Knutson}, {Lecavelier Des
  Etangs}, {Pont}, {Showman}, {Vidal-Madjar}, {Williamson}, \&
  {Wilson}}]{Sing2016}
{Sing}, D.~K., {Fortney}, J.~J., {Nikolov}, N., {et~al.} 2016, \nat, 529, 59

\bibitem[{{Siverd} {et~al.}(2012){Siverd}, {Beatty}, {Pepper}, {Eastman},
  {Collins}, {Bieryla}, {Latham}, {Buchhave}, {Jensen}, {Crepp}, {Street},
  {Stassun}, {Gaudi}, {Berlind}, {Calkins}, {DePoy}, {Esquerdo}, {Fulton},
  {F{\H{u}}r{\'e}sz}, {Geary}, {Gould}, {Hebb}, {Kielkopf}, {Marshall},
  {Pogge}, {Stanek}, {Stefanik}, {Szentgyorgyi}, {Trueblood}, {Trueblood},
  {Stutz}, \& {van Saders}}]{Siverd2012}
{Siverd}, R.~J., {Beatty}, T.~G., {Pepper}, J., {et~al.} 2012, \apj, 761, 123

\bibitem[{{Spiegel} {et~al.}(2009){Spiegel}, {Silverio}, \&
  {Burrows}}]{Spiegel2009}
{Spiegel}, D.~S., {Silverio}, K., \& {Burrows}, A. 2009, \apj, 699, 1487

\bibitem[{{Stassun} {et~al.}(2017){Stassun}, {Collins}, \&
  {Gaudi}}]{Stassun:2017}
{Stassun}, K.~G., {Collins}, K.~A., \& {Gaudi}, B.~S. 2017, \aj, 153, 136

\bibitem[{{Stassun} \& {Torres}(2016)}]{Stassun:2016}
{Stassun}, K.~G. \& {Torres}, G. 2016, \aj, 152, 180

\bibitem[{{Stassun} \& {Torres}(2018)}]{Stassun:2018}
{Stassun}, K.~G. \& {Torres}, G. 2018, \apj, 862, 61

\bibitem[{{Stellingwerf}(1978)}]{Stellingwerf1978}
{Stellingwerf}, R.~F. 1978, \apj, 224, 953

\bibitem[{{Tan} \& {Komacek}(2019)}]{Tan2019}
{Tan}, X. \& {Komacek}, T.~D. 2019, \apj, 886, 26

\bibitem[{{Torres} {et~al.}(2010){Torres}, {Andersen}, \&
  {Gim{\'e}nez}}]{Torres:2010}
{Torres}, G., {Andersen}, J., \& {Gim{\'e}nez}, A. 2010, \aapr, 18, 67

\bibitem[{{Tsiaras} {et~al.}(2019){Tsiaras}, {Waldmann}, {Tinetti}, {Tennyson},
  \& {Yurchenko}}]{Tsiaras2019}
{Tsiaras}, A., {Waldmann}, I.~P., {Tinetti}, G., {Tennyson}, J., \&
  {Yurchenko}, S.~N. 2019, Nature Astronomy, 3, 1086

\bibitem[{{von Essen} {et~al.}(2017){von Essen}, {Cellone}, {Mallonn},
  {Albrecht}, {Micul{\'a}n}, \& {M{\"u}ller}}]{vonessen2017}
{von Essen}, C., {Cellone}, S., {Mallonn}, M., {et~al.} 2017, \aap, 603, A20

\bibitem[{{von Essen} {et~al.}(2020{\natexlab{a}}){von Essen}, {Mallonn},
  {Borre}, {Antoci}, {Stassun}, {Khalafinejad}, \&
  {Tautvai{\v{s}}ien{\.{e}}}}]{vonEssen2020b}
{von Essen}, C., {Mallonn}, M., {Borre}, C.~C., {et~al.} 2020{\natexlab{a}},
  \aap, 639, A34

\bibitem[{{von Essen} {et~al.}(2020{\natexlab{b}}){von Essen}, {Mallonn},
  {Hermansen}, {Nixon}, {Madhusudhan}, {Kjeldsen}, \&
  {Tautvai{\v{s}}ien{\.{e}}}}]{vonEssen2020}
{von Essen}, C., {Mallonn}, M., {Hermansen}, S., {et~al.} 2020{\natexlab{b}},
  \aap, 637, A76

\bibitem[{{von Essen} {et~al.}(2019){von Essen}, {Mallonn}, {Welbanks},
  {Madhusudhan}, {Pinhas}, {Bouy}, \& {Weis Hansen}}]{vonEssen2019AlO}
{von Essen}, C., {Mallonn}, M., {Welbanks}, L., {et~al.} 2019, \aap, 622, A71

\bibitem[{{von Essen} {et~al.}(2013){von Essen}, {Schr{\"o}ter}, {Agol}, \&
  {Schmitt}}]{vonEssen2013}
{von Essen}, C., {Schr{\"o}ter}, S., {Agol}, E., \& {Schmitt}, J.~H.~M.~M.
  2013, \aap, 555, A92

\bibitem[{Wang {et~al.}(2019)Wang, Jones, Shporer, Fulton, Paredes, Trifonov,
  Kossakowski, Eastman, Redfield, G\"{u}nther, Kreidberg, Huang, Millholland,
  Seligman, Fischer, Brahm, Wang, Cruz, Henry, James, Addison, Liang, Davis,
  Tronsgaard, Worku, Brewer, K\"{u}rster, Zhang, Beichman, Bieryla, Brown,
  Christiansen, Ciardi, Collins, Esquerdo, Howard, Isaacson, Latham, Mazeh,
  Petigura, Quinn, Shahaf, Siverd, Rodler, Reffert, Zakhozhay, Ricker,
  Vanderspek, Seager, Winn, Jenkins, Boyd, F{\H{u}}r{\'{e}}sz, Henze, Levine,
  Morris, Paegert, Stassun, Ting, Vezie, \& Laughlin}]{Wang2019}
Wang, S., Jones, M., Shporer, A., {et~al.} 2019, The Astronomical Journal, 157,
  51

\bibitem[{{Welsh} {et~al.}(2010){Welsh}, {Orosz}, {Seager}, {Fortney},
  {Jenkins}, {Rowe}, {Koch}, \& {Borucki}}]{Welsh2010}
{Welsh}, W.~F., {Orosz}, J.~A., {Seager}, S., {et~al.} 2010, \apjl, 713, L145

\bibitem[{White {et~al.}(1958)White, Johnson, \& Dantzig}]{White1958}
White, W.~B., Johnson, S.~M., \& Dantzig, G.~B. 1958, The Journal of Chemical
  Physics, 28, 751

\bibitem[{{Winn} {et~al.}(2008){Winn}, {Holman}, {Torres}, {McCullough},
  {Johns-Krull}, {Latham}, {Shporer}, {Mazeh}, {Garcia-Melendo}, {Foote},
  {Esquerdo}, \& {Everett}}]{Winn2008}
{Winn}, J.~N., {Holman}, M.~J., {Torres}, G., {et~al.} 2008, \apj, 683, 1076

\bibitem[{{Wong} {et~al.}(2020{\natexlab{a}}){Wong}, {Benneke}, {Shporer},
  {Fetherolf}, {Kane}, {Ricker}, {Vand erspek}, {Seager}, {Winn}, {Collins},
  {Mireles}, {Morris}, {Tenenbaum}, {Ting}, {Rinehart}, \&
  {Villase{\~n}or}}]{Wong2020}
{Wong}, I., {Benneke}, B., {Shporer}, A., {et~al.} 2020{\natexlab{a}}, \aj,
  159, 104

\bibitem[{{Wong} {et~al.}(2020{\natexlab{b}}){Wong}, {Shporer}, {Daylan},
  {Benneke}, {Fetherolf}, {Kane}, {Ricker}, {Vanderspek}, {Latham}, {Winn},
  {Jenkins}, {Boyd}, {Glidden}, {Goeke}, {Sha}, {Ting}, \&
  {Yahalomi}}]{Wong2020b}
{Wong}, I., {Shporer}, A., {Daylan}, T., {et~al.} 2020{\natexlab{b}}, \aj, 160,
  155

\bibitem[{{Wong} {et~al.}(2020{\natexlab{c}}){Wong}, {Shporer}, {Kitzmann},
  {Morris}, {Heng}, {Hoeijmakers}, {Demory}, {Ahlers}, {Mansfield}, {Bean},
  {Daylan}, {Fetherolf}, {Rodriguez}, {Benneke}, {Ricker}, {Latham},
  {Vanderspek}, {Seager}, {Winn}, {Jenkins}, {Burke}, {Christiansen}, {Essack},
  {Rose}, {Smith}, {Tenenbaum}, \& {Yahalomi}}]{Wong2019b}
{Wong}, I., {Shporer}, A., {Kitzmann}, D., {et~al.} 2020{\natexlab{c}}, \aj,
  160, 88

\bibitem[{{Yurchenko} {et~al.}(2011){Yurchenko}, {Barber}, \&
  {Tennyson}}]{Yurchenko2011}
{Yurchenko}, S.~N., {Barber}, R.~J., \& {Tennyson}, J. 2011, \mnras, 413, 1828

\bibitem[{{Yurchenko} \& {Tennyson}(2014)}]{Yurchenko2014a}
{Yurchenko}, S.~N. \& {Tennyson}, J. 2014, \mnras, 440, 1649

\bibitem[{{Yurchenko} {et~al.}(2013){Yurchenko}, {Tennyson}, {Barber}, \&
  {Thiel}}]{Yurchenko2013}
{Yurchenko}, S.~N., {Tennyson}, J., {Barber}, R.~J., \& {Thiel}, W. 2013,
  Journal of Molecular Spectroscopy, 291, 69

\bibitem[{{Zechmeister} \& {K{\"u}rster}(2009)}]{LombScargle}
{Zechmeister}, M. \& {K{\"u}rster}, M. 2009, \aap, 496, 577

\bibitem[{{Zhang} {et~al.}(2018){Zhang}, {Knutson}, {Kataria}, {Schwartz},
  {Cowan}, {Showman}, {Burrows}, {Fortney}, {Todorov}, {Desert}, {Agol}, \&
  {Deming}}]{Zhang2018}
{Zhang}, M., {Knutson}, H.~A., {Kataria}, T., {et~al.} 2018, \aj, 155, 83

\end{thebibliography}

\begin{appendix}

\section{Posterior distributions}

\begin{figure}[ht!]
    \centering
    \includegraphics[width=0.5\textwidth]{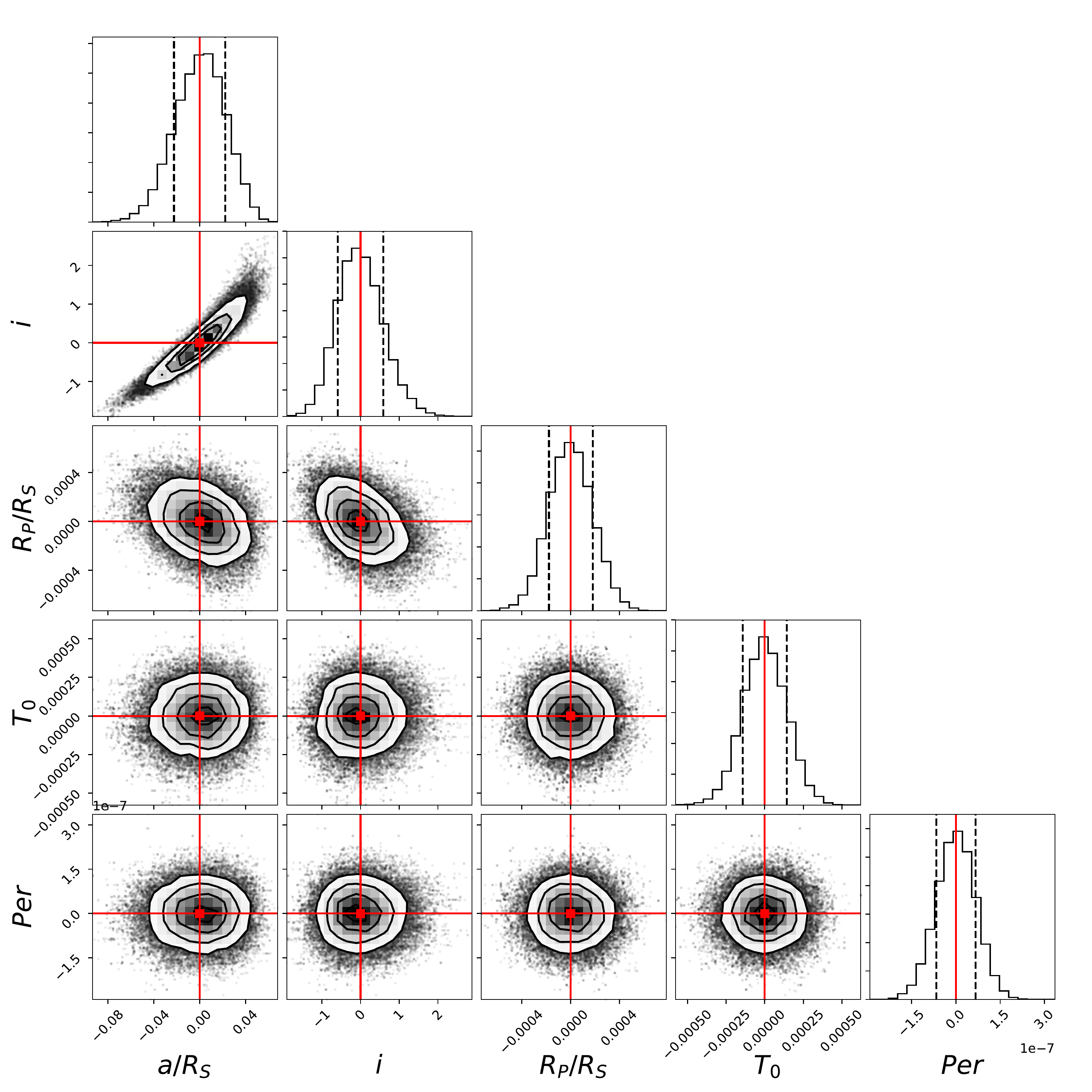}
    \caption{\label{fig:posterior_transits} Posterior distributions for the primary transit parameters assuming a circular orbit for KELT-1b. In the figure, the red points indicate the best-fit values, and the colors of the contours highlight the 1, 2 and 3-$\sigma$ intervals. To allow for a better visual inspection of the uncertainties, the posteriors were shifted to the best-fit values specified in Table~\ref{tab:transit_parameters}.}
\end{figure}

\begin{figure*}[ht!]
    \centering
    \includegraphics[width=.99\textwidth]{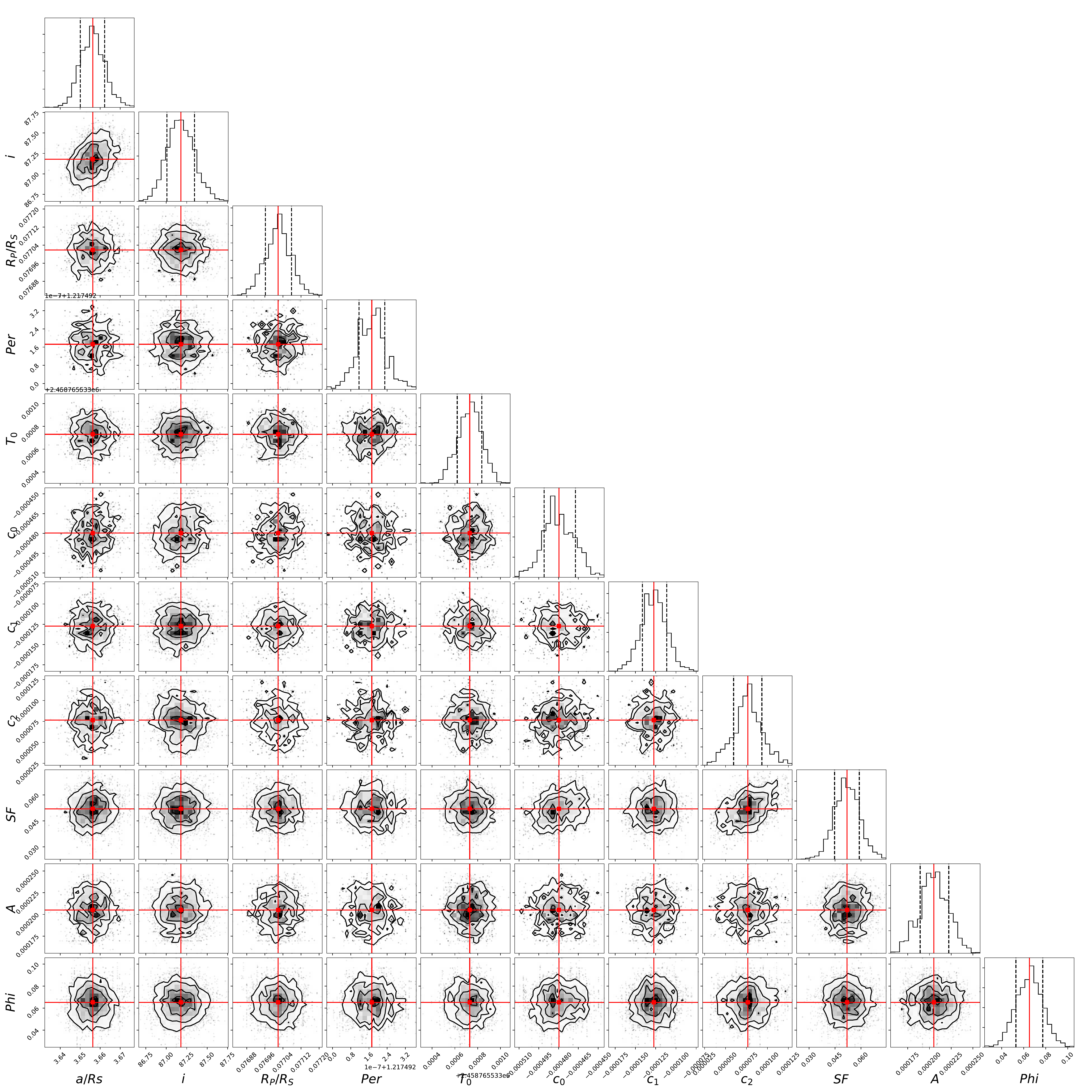}
    \caption{\label{fig:posteriors_M1} Posterior distributions for the eleven parameters determined from our simultaneous fit, M1.}
\end{figure*}

\begin{figure*}[ht!]
    \centering
    \includegraphics[width=.99\textwidth]{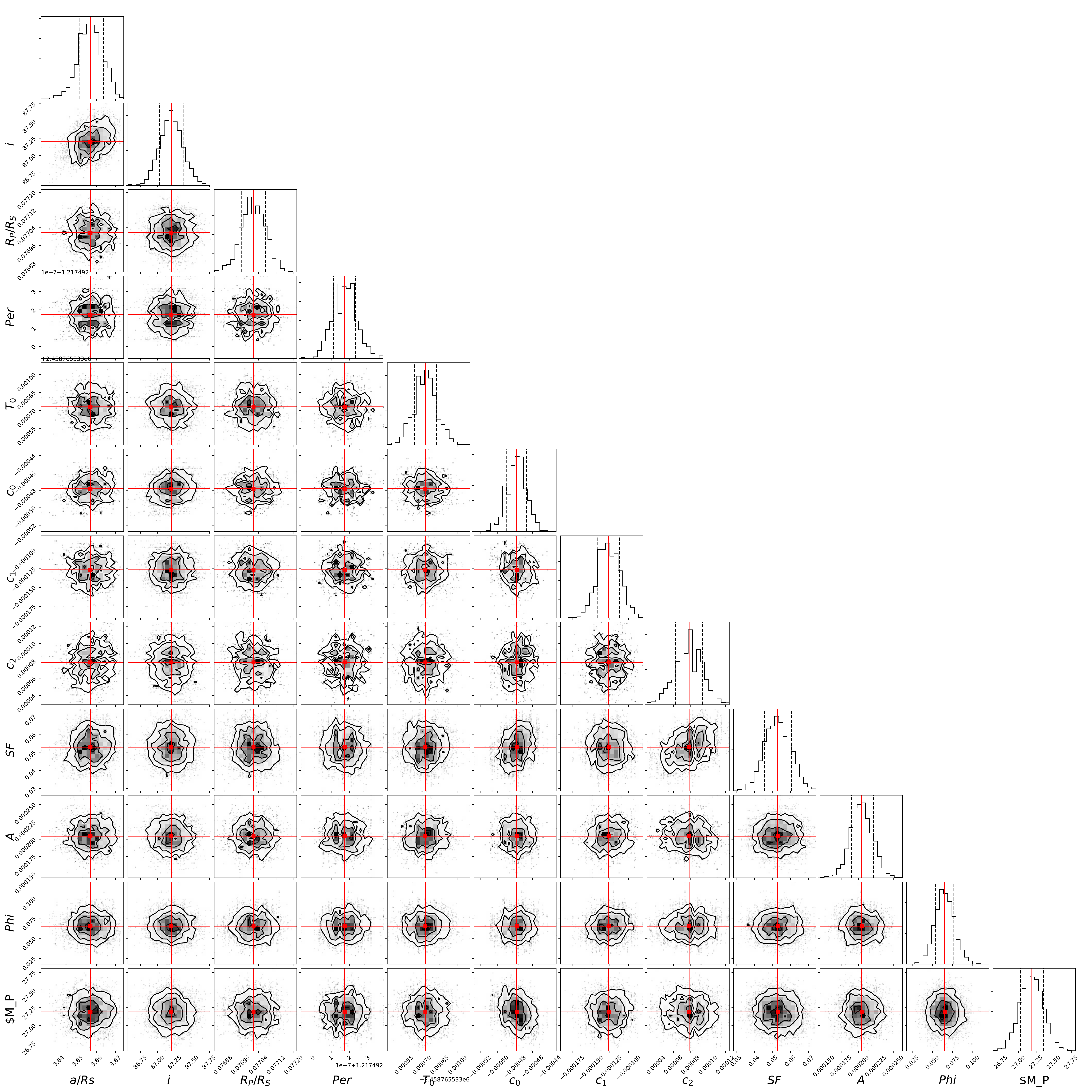}
    \caption{\label{fig:posteriors_M2} Posterior distributions for the twelve parameters determined from our simultaneous fit, M3.}
\end{figure*}

\end{appendix}

\end{document}